\documentclass[a4paper,11pt]{article}

\pdfoutput=1 % if your are submitting a pdflatex (i.e. if you have
\usepackage{jheppub}

\usepackage[T1]{fontenc} % if needed

\def\theequation{\thesection.\arabic{equation}}

\def\be{\begin{equation}}
\def\ee{\end{equation}}
\def\ba{\begin{eqnarray}}
\def\ea{\end{eqnarray}}
\def\lb{\label}
\def\nn{\nonumber}

\def\a{\alpha}
\def\b{\beta}

\def\d{\delta}

\def\e{\varepsilon}
\def\l{\lambda}

\def\s{\sigma}

\def\G{\Gamma}

\def\T{\Theta}

\def\Z{\mathbb Z}

\def\id{\mbox{\em 1\hspace{-3.4pt}I}}

\def\uz{\underline{z}}

\usepackage[dvipsnames]{xcolor}
\definecolor{deepgreen}{HTML}{009504}

\title{\boldmath Braiding Fibonacci anyons}

\author{Ludmil Hadjiivanov}
\author{and Lachezar S. Georgiev}
\affiliation{Institute for Nuclear Research and Nuclear Energy, \\ Bulgarian Academy of Sciences ,\\
Tsarigradsko Chaussee 72, 1784 Sofia, Bulgaria}

\emailAdd{lhadji@inrne.bas.bg}
\emailAdd{lgeorg@inrne.bas.bg}

\abstract{Fibonacci anyons $\e\,$ provide the simplest possible model of non-Abelian fusion rules: $[1] \times [1] = [0] \oplus [1]\,.$ We propose a conformal field theory construction of topological quantum registers based on Fibonacci anyons realized as quasiparticle excitations in the $\Z_3\,$ parafermion fractional quantum Hall state. To this end, the results of Ardonne and Schoutens %\cite{AS07} 
for the correlation function of four Fibonacci fields are extended to the case of arbitrary number $n\,$ of quasi-holes and $N=3\, r\,$ electrons. Special attention is paid to the braiding properties of the obtained correlators. We explain in details the construction of a monodromy representation of the Artin braid group ${\cal B}_n\,$ acting on $n$-point conformal blocks of Fibonacci anyons. The matrices of braid group generators are displayed explicitly for all $n\le 8\,.$ A simple recursion formula makes it possible to extend without efforts the construction to any $n\,.$ Finally, we construct ${\cal N}\,$ qubit computational spaces in terms of conformal blocks of $2\, {\cal N} + 2\,$ Fibonacci anyons. 
}


\begin{document} 

\maketitle

\flushbottom

\section{Introduction}

\setcounter{equation}{0}
\renewcommand\theequation{\thesection.\arabic{equation}}

Noise and decoherence are basic challenges to quantum computation. The limitation of quantum data vulnerability by fundamental physical principles is therefore highly welcomed. Such theoretical possibility is provided by Topological Quantum Computation (TQC) (cf. e.g. the recent textbook \cite{S23}). An experimentally promising setting of this type in which quantum information processing is protected from noise and decoherence by the topological properties of the quantum system is based on braiding of non-Abelian anyons realized as quasiparticle excitations in certain fractional quantum Hall (FQH) states. We will focus in what follows on the Fibonacci anyons realized as the $\e\,$ field in the $\Z_3\,$ parafermion FQH state in the second Landau level with filling factor $\frac{12}{5}\,.$ This construction is introduced and discussed in details in the parallel paper \cite{GHM24} where the reader could find as well an extensive list of literature on the subject.

The aim of present article is to provide a description of the {\em monodromy representation} of the braid group ${\cal B}_n\,$ (see e.g. \cite{TH01} and the references therein) realized on the set of chiral conformal blocks of $n\,$ Fibonacci anyons $\e\,$ and $3\,r\,$ electrons. To this end we will follow the Ardonne and Schoutens' approach \cite{AS07} which we will generalize in two different directions. While \cite{AS07} only covers the case of four ($n=4\,$)  $\e\,$ fields  and no electrons ($r=0\,$), we will consider also arbitrary $r\,$ case and then the most general case thus opening a theoretical perspective to constructing ${\cal N}$ qubit quantum registers.

We will also pay the due attention to the properties of the obtained $n$-point multivalued correlation functions with respect to the exchange of neighboring fields along certain (homotopy classes of) paths. These "half monodromy" transformations define the generators of the braid group ${\cal B}_n\,$ and hence, must satisfy the corresponding Artin relations. The proper treatment of the subject requires the introduction of dual bases related by the so called fusion matrices, cf. e.g. John Preskill's classical lectures \cite{P04} or Steven Simon's modern textbook \cite{S23}. The basic, four point ($n=4\,$) case is dealt with care and the obtained results are generalized to arbitrary number $n\,$ of the Fibonacci fields. A recursive formula for the matrices of the ${\cal B}_n\,$ generators in a suitable basis of conformal blocks is given which also allows, in principle, an easy verification of the Artin relations.

The organization of the paper is the following. In Section 2 we perform a derivation of the conformal blocks of four Fibonacci anyons in full detail reproducing the results obtained in \cite{AS07}. We shed light on some crucial points and introduce our own conventions. We conclude this section with the calculation of the (diagonal) matrices of the generators $b_1\,$ and $b_3\,$ in the corresponding two dimensional representation of the braid group ${\cal B}_4\,.$ In Section 3 we define the dual basis and the fusion matrix $F\,,$ calculate the matrix of $b_2\,$ and verify the Artin triple relations. In Section 4 we generalize the obtained results to the case when electrons are also present (due to the $\Z_3\,$ symmetry, the functions are only nonzero when their number is a multiple of three), applying a slightly different technique to compute the braid matrices. In Section 5 we construct a basis in the space of Fibonacci conformal blocks for arbitrary $n\,$ and propose an algorithm for finding the braid group action on it. The results are made explicit in Section 6, for $n=5\,,\, 6\,,\, 7 \,$ and $8\,.$ A recursive construction of the braid group action in the general case is introduced in Section 7. In the concluding Section 8 we construct ${\cal N}\,$ qubit computational spaces in terms of conformal blocks of $2\, {\cal N} + 2\,$ Fibonacci anyons. 

There is a vast literature on Fibonacci anyons and their possible use in TQC. On top of the fundamental papers by Preskill and Ardonne-Schoutens mentioned above we have taken inspiration from the work of N. Bonesteel, L. Hormozi, G. Zikos and S. Simon \cite{BHZS05, HZBS07} and from others. Albeit having some technical overlap with the present article, \cite{GHL22} that appeared in the course of our investigation is quite different in scope from it.

TQC based on non-Abelian and, in particular, Fibonacci anyons has been subject of intensive study already for more than a quarter of a century. From a broader perspective, one can ask if the conformal field theory (CFT) methods based on the coordinate representation of the anyon wave function in terms of CFT correlators are more efficient than some abstract computational methods of algebraic, e.g. quantum group \cite{SB01}, or categorical \cite{FKLW02, FLW02} type focused on topological rather than dynamical aspects of the relevant low dimensional physical systems. In this respect we would agree with the authors of \cite{SB01} that each of these approaches provides a useful way of thinking complementary to the other.

To conclude, we will summarize the most important results of the present paper as we see them:

- the Ardonne and Schoutens' analytic result for $n=4\,$ Fibonacci anyons ($\e\,$ fields)  \cite{AS07} has been generalized to any $n\,$ and any (presumably, great) number $3 r\,$ of electrons and the basic Fibonacci $R\,$ and $F\,$ matrices have been derived in the general case by analytic continuation. This is a new theoretical result which could be of practical use as well (see \cite{GHM24}). 

- the braid generating matrices in the natural basis of conformal blocks have been found explicitly for any $n\,$ (and proved to be independent of $r\,$) by following a new, algorithmic and efficient prescription based on the natural chain of inclusions ${\cal B}_{n-2} \subset {\cal B}_{n-1} \subset {\cal B}_n\,.$

- ${\cal N}\,$ qubit computational subspaces of the space of $n = 2\, {\cal N} + 2\,$ anyon conformal blocks have been identified in this setting\footnote{On a more abstract level, qubit encoding by pairs of Fibonacci anyons has been considered already by Freedman et al. \cite{FKLW02, FLW02}. The idea to devise qubit encoding in terms of Fibonacci anyon pairs is actually quite natural as, due to the fusion rules, the latter form effectively two level quantum systems.} for any ${\cal N}\,.$

\section{Conformal blocks of four Fibonacci anyons ($n=4$)}
%%\,,\ r=0$)}

\setcounter{equation}{0}
\renewcommand\theequation{\thesection.\arabic{equation}}

As shown in \cite{GHM24}, it is physically plausible to present the coordinate wave function of 4 Fibonacci anyons and $N = 3\, r\,$ electron holes in the plane, up to a non-holomorphic Gaussian exponential factor, into the following split form containing a $\Z_3\,$ parafermion (PF) part and another, Abelian ${\widehat{u(1)}}\,$ one of Laughlin type:
\ba
&&\Psi_{4 F} (w_1, \dots, w_4; \, z_1,\dots , z_{3r}) = 
\langle\, \e (w_1) \dots \e (w_4) \,\prod_{i=1}^{3r} \psi_1 (z_i)\, \rangle_{PF}\times\nn\\
&&\times\, \prod_{1\le a< b\le 4} (w_a - w_b)^{\frac{3}{5}} \,
\prod_{a=1}^4 \prod_{i=1}^N (w_a-z_i) \,\prod_{1\le j <\le \ell} (z_j-z_\ell)^{\frac{5}{3}}\ .
\lb{4eps}
\ea
The complex coordinates $w_a\,,\ a=1,2,3,4\,$ and $z_i \,,\ i = 1,\dots , 3\,r\,$ correspond to the positions of the anyons and the electron holes, respectively. 

The Abelian current algebra (extended with a conjugate pair of appropriate vertex exponents) plays an important role, as it allows the model to incorporate the electrically charged edge excitations in the FQH liquid. However, we will be interested exclusively in this paper in the exchange (braiding) properties of the neighboring Fibonacci fields, and from this point of view the contribution of the Abelian part in (\ref{4eps}) is trivial. For this reason we will concentrate in what follows on the parafermionic part only. This will lead to the lack of an overall factor $e^{i\frac{3\pi}{5}}\,$ ( $= q^3\,,$ see below) in the braiding matrices derived in the present paper with respect to those in \cite{GHM24}; the Artin braid relations are indifferent to this change.

Following the procedure well described in \cite{AS07}, the correlation functions
\be
\langle\, \e (w_1) \dots \e (w_4) \,\prod_{j=5}^{3r+4}\psi_1 (z_j)\, \rangle^{(p)}\ , \quad p=0,1\,
\lb{4epsAS}
\ee
of four Fibonacci anyons and $3\,r\,$ "electrons"\footnote{A somewhat loosely attributed name, for short, which generalizes a special case ($M=1\,$ in (\ref {Prgen})).} can be obtained by fusing $\Z_3\,$ parafermionic fields $\s_1\,$ and $\psi_1\,$ with arguments at $\{ w_2, w_1, w_3, w_4 \}\,$ and $\{ z_1, z_2, z_3, z_4 \}\,,$ respectively, subject to the operator product expansion (OPE)
\be
\s_1 (w)\, \psi_1 (z) = \sqrt{\frac{2}{3}}\, (w-z)^{- \frac{1}{3}}\, \e (z) \, +\, o (w-z)\ ,\qquad{\rm or}\qquad \s_1 \, \psi_1 = \sqrt{\frac{2}{3}} \,\, \e
\lb{OPEspe}
\ee
for short, so that
\ba
&&\langle\, \e (w_1) \dots \e (w_4) \,\prod_{j=5}^{3r+4}\psi_1 (z_j)\, \rangle^{(p)} =\nn\\
&&= \ \frac{9}{4}\,\lim\limits_{\{z\} \to \{w\}}\,\prod_{j=1}^4 (w_j-z_j)^{\frac{1}{3}}\,\langle \, \s_1(w_1) \dots \s_1(w_4)\, \psi_1(z_1) \dots \psi_1 (z_4)\,\prod_{j=5}^{3r+4}\psi_1 (z_j) \rangle^{(p)}\ . \qquad
\lb{4efusion}
\ea
The chiral fields in (\ref{OPEspe}) with conformal dimensions
\be
\Delta_{\s_1} = \frac{1}{15}\ ,\quad
\Delta_{\psi_1} = \frac{2}{3}\ ,\quad
\Delta_{\e} = \frac{2}{5}
\lb{dim}
\ee
are special cases ($k = 3\,$ and $\e \equiv \e^{(1)}$) of the general setting in \cite{ZF85} for $\Z_k\,$ parafermions where
\ba
&&\Delta_{\s_\ell} = \frac{\ell (k-\ell) }{2\, k (k+2)}\ ,\quad
\Delta_{\psi_\ell} = \frac{\ell (k-\ell )}{k}\ ,\quad \ell = 1, 2,\dots k-1\ ,\nn\\
&&\Delta_{\e^{(j)}} = \frac{j (j+1)}{k+2}\ ,\quad j = 1, \dots , [k/2]
\lb{dimZF}
\ea
are the dimensions of the order parameters $\s_\ell\,,$ the parafermionic currents $\psi_\ell\,,$ and the $\Z_k$-neutral fields $\e^{(j)}\,,$ respectively. The fusion of two $\s_1\,$ reads
\be
\s_1\,\s_1 = \frac{1}{\sqrt{3}} \,\psi_1 + \sqrt{2\,C}\, \s_2\ ,\qquad 2\, C = \sqrt{\frac{\Gamma(\frac{1}{5}) \Gamma^3 (\frac{3}{5})}{ \Gamma (\frac{4}{5}) \Gamma^3 (\frac{2}{5})}}
\lb{ss}
\ee
and the two fusion channels $(0)\,$ and $(1)\,$ for the parafermion correlator correspond to $\s_1 \,\s_1 \,\sim  \, \psi_1\,$ and $\s_1 \,\s_1 \,\sim  \, \s_2\,,$ respectively. Eq.(3.3) in \cite{AS07}  is equivalent in the case under consideration to
\ba
&&\langle \s_1(w_1) \dots \s_1(w_4)\, \psi_1(z_1) \dots \psi_1 (z_4)
\,\prod_{j=5}^{3r+4}\psi_1 (z_j) \, \rangle^{(p)} = \lb{spP}\\
&&= P (\{ w \},\{ z \} )  \
\Psi_{RR}^{(0,1)} (w_1, w_2, w_3, w_4 ; z_1, z_2, z_3, z_4, z_5, \dots , z_{3r+4} )\ .
\nn
\ea
Here the prefactor is given by
\be
P (\{ w \},\{ z \} ) = \prod_{1\le i < j \le 3r+4}z_{ij}^{-\frac{2}{3} - M}\, \prod_{i=1}^4 \prod_{j=1}^{3r+4} (w_i-z_j)^{-\frac{1}{3}} \,\prod_{1\le i<j \le 4} w_{ij}^{-\frac{1}{6} + \frac{M}{2(3M+2)} }
\lb{Prgen}
\ee
where $z_{ij} = z_i - z_j\,,\ w_{ij} = w_i - w_j\,,$
and the $k=3\,$ Read-Rezayi (RR) wave functions $\Psi_{RR}^{(p)}\,,\ p= 0,1\,$ of $n=4\,$ quasi-holes and  $N = 3\, r + 4\,$  electrons are expressed as
\ba
&&\Psi_{RR}^{(p)} (w_1, w_2, w_3, w_4 ; z_1, z_2, z_3, z_4, z_5, \dots , z_{3r+4}) = \lb{AB}\\
&& = A^{(p)} (\{ w \})\, \Psi_{12,34} (\{ w \} ; \{ z \}) +
B^{(p)} (\{ w \})\, \Psi_{13,24} (\{ w \} ; \{ z \}) \ ,\quad p= 0,1\ ,
\nn
\ea
cf. \cite{NW96, CGT01}. The coefficient functions $A^{(p)} (\{ w \})\,$ and $B^{(p)} (\{ w \})\,$  calculated in \cite{AS07} %(see (3.12) therein)
are given by
\ba
&&A^{(0)} (\{ w \}) = (w_{12} w_{34})^{\frac{7}{10}}\, x^{\frac{3}{10}}\, {\cal F}^{(0)}_1 (x)\ ,\nn\\
&&B^{(0)} (\{ w \}) = - (w_{12} w_{34})^{\frac{7}{10}}\, x^{-\frac{7}{10}}\, (1-x)\, {\cal F}^{(0)}_2 (x)\ , \nn\\
&&A^{(1)} (\{ w \}) = - (-1)^{\frac{2}{5}} \, C\, (w_{12} w_{34})^{\frac{7}{10}}\, x^{\frac{3}{10}}\, {\cal F}^{(1)}_1 (x) \ ,\nn\\
&&B^{(1)} (\{ w \}) = (-1)^{\frac{2}{5}}\, C \, (w_{12} w_{34})^{\frac{7}{10}}\, x^{-\frac{7}{10}}\, (1-x)\, {\cal F}^{(1)}_2 (x)\ ,
\lb{AB1}
\ea
the harmonic ratio $x\,$ being defined as
\be
x = \frac{w_{12} w_{34}}{w_{14}w_{32}} %%\ ,\quad w_{ij}:= w_i-w_j\ ,
\lb{x}
\ee
and the functions ${\cal F}^{(p)}_{1,2}(x)\,$ %(borrowed from \cite{KZ84})
are expressed in terms of hypergeometric functions %\cite{BE53}
as follows:
%(see (A.12) in \cite{AS07}):
\ba
&&{\cal F}^{(0)}_1 (x) = x^{-\frac{3}{10}}\, (1-x)^{\frac{1}{10}}\,
F(\frac{1}{5} , -\frac{1}{5} , \frac{3}{5} ; x )\ ,\nn\\
&&{\cal F}^{(0)}_2 (x) = \frac{1}{3}\, x^{\frac{7}{10}}\, (1-x)^{\frac{1}{10}}\,F(\frac{6}{5} , \frac{4}{5} , \frac{8}{5} ; x )\ ,\nn\\
&&{\cal F}^{(1)}_1 (x) = x^{\frac{1}{10}}\, (1-x)^{\frac{1}{10}}\,F(\frac{1}{5} , \frac{3}{5} , \frac{7}{5} ; x )\ ,\nn\\
&&{\cal F}^{(1)}_2 (x) = - 2\,x^{\frac{1}{10}}\, (1-x)^{\frac{1}{10}}\,F(\frac{1}{5} , \frac{3}{5} , \frac{2}{5} ; x )\ .
\lb{HGF}
\ea
The factors $\Psi_{12,34} (\{ w \} ; \{ z \})\,$
and $\Psi_{13,24} (\{ w \} ; \{ z \})\,$ in (\ref{AB}) correspond to two of the possible splittings of the four quasi-holes into two pairs so that, in principle, there is one more possibility which, however, does not produce an independent function \cite{NW96}, as
\be
\Psi_{14,23} (\{ w \} ; \{ z \}) = x\, \Psi_{12,34} (\{ w \} ; \{ z \}) + (1-x) \, \Psi_{13,24} (\{ w \} ; \{ z \})\ .
\lb{NWx}
\ee
%(see (3.6) in \cite{AS07}).
One divides the $N = 3\, r +4 \,$ electrons into three groups, $S_1\,$ containing $r+2\,$ electrons, and $S_2\,$ and $S_3\,,$ containing $r+1\,$ electrons each. The two remaining factors are homogeneous polynomials in the differences of the quasi-hole and electron coordinates of the kind
%(cf. (3.5), (3.1) in \cite{AS07})
\ba
&&\Psi_{12,34} (\{ w \} ; \{ z \}) = \frac{3^{-\frac{3}{2} r} }{9} \times
\nn\\
&&\times \,\sum_{S_1, S_2, S_3}\, \prod_{{i \in S_2}} (z_i - w_1) (z_i - w_2)\, \prod_{j \in S_3} (z_j - w_3) (z_j - w_4)\, \Psi_{S_1}^2 (\{ z \})\, \Psi_{S_2}^2 (\{ z \})\, \Psi_{S_3}^2 (\{ z \})\ ,\nn\\
&&\Psi_{13,24} (\{ w \} ; \{ z \}) = \frac{3^{-\frac{3}{2} r} }{9} \times \lb{PSI}
\\
&&\times \,\sum_{S_1, S_2, S_3}\, \prod_{{i \in S_2}} (z_i - w_1) (z_i - w_3)\, \prod_{j \in S_3} (z_j - w_2) (z_j - w_4)\, \Psi_{S_1}^2 (\{ z \})\, \Psi_{S_2}^2 (\{ z \})\, \Psi_{S_3}^2 (\{ z \})\ ,\nn
\ea
where
\be
\Psi_{S_i}^2 (\{ z \}) = \prod_{{i<j}\atop{i,j \in S_i}} z_{ij}^2\ ,\qquad i=1,2,3\ .
\lb{Laughlin}
\ee
Obviously, each of the sums in (\ref{PSI}) contains 
\begin{equation*}
\binom{3r+4}{r+2\,,\, r+1\,,\, r+1} = \frac{(3r+4)!}{(r+2)!\,(r+1)!\,(r+1)!}\,
\end{equation*} 
terms (corresponding to the various options of attributing $r+2\,$ of the $3r+4\,$ electrons to the group $S_1\,$ times the $\binom{2r+2}{r+1}\,$ different choices to evenly distribute the remaining $2r+2\,$ ones between $S_2\,$ and $S_3$). The corresponding dimension in "mass" (inverse length) units is equal to minus the overall order of the homogeneous polynomials (\ref{PSI}):
\be
\Delta_\Psi = - 3(r+1)(r+2) = - 3\, r^2 - 9\, r - 6\,.
\lb{DPsi}
\ee

\medskip

Due to the identity $w_{12}w_{34}=w_{14}w_{32}+w_{13}w_{24}\,,$ one has
\be
x = \frac{w_{12} w_{34}}{w_{14}w_{32}}\qquad\Rightarrow\qquad
1-x = \frac{w_{13} w_{42}}{w_{14}w_{32}}\ ,\qquad \frac{x}{1-x} = \frac{w_{12} w_{34}}{w_{13}w_{42}}\ .
\lb{x1}
\ee
We will use in what follows an alternative harmonic ratio as well,
\ba
&&\eta = \frac{w_{12} w_{34}}{w_{13} w_{24}} = \frac{x}{x-1}\quad {\rm so\ that}\quad x = \frac{\eta}{\eta -1}\ , \quad 1-x = \frac{1}{1-\eta} \quad {\rm and}\nn\\
&&w_{13} w_{24} = \frac{1}{\eta}\,w_{12} w_{34}\ ,\quad w_{14} w_{23} = \frac{1-\eta}{\eta}\,w_{12} w_{34}\ .
\lb{eta-x}
\ea
Note that for $w_i\,$ real and naturally ordered, $w_1 > w_2 > w_3 > w_4\,,$ one has $0 < \eta < 1\,.$

\medskip

We will first compute the coefficient of $\prod_{i=1}^4 (w_i-z_i)^{-\frac{1}{3}}\,$ arising in the limit $z_i \to w_i\,,\ i= 1,\dots ,4\,$ of the prefactor $P (\{ w \},\{ z \} )\,$ (\ref{Prgen}) of overall dimension (in mass units)
\be
\Delta_P = 3\, r^2 + 11\, r + \frac{31}{3} + M \left[ \binom{3r+4}{2} -\frac{3}{3M+2}\, \right]\ .
\lb{DPr}
\ee
The result is
\ba
&& \lim\limits_{z_i \to w_i}\, \prod_{i=1}^4 (w_i-z_i)^{\frac{1}{3}}\,
P (\{ w \},\{ z \} ) = \nn\\
&&= \prod_{1\le i<j \le 4} w_{ij}^{-3 \frac{(M+1)^2}{3M+2} } \prod_{i=1}^4 \prod_{j=5}^{3r+4} (w_i-z_j)^{-(M+1)} \,
\prod_{5\le i < j \le 3r+4}z_{ij}^{-(M+\frac{2}{3})}\ .\qquad\quad
\lb{PrMr}
\ea
We will assume from now on that $M = 0\,.$ Combining (\ref{DPr}) with (\ref{DPsi}) and the $- \frac{7}{5}\,$ factor coming from (\ref{AB1}), we can easily find that their sum (for $M=0$) reproduces the dimension of the parafermionic correlator in (\ref{spP}),
\be
3\, r^2 + 11\, r + \frac{31}{3} - 3\, r^2 - 9\, r - 6 - \frac{7}{5} \ (\, = 2\, r + \frac{44}{15}\,)\, = \frac{4}{15} + \frac{2}{3}\,(3\,r+4) \ ,
\lb{sumD}
\ee
thus verifying an obvious consistency condition implied by scale invariance (the invariance with respect to $\phi (z) \to \l^{\Delta_\phi}\, \phi (\l\, z)$ for all conformal fields).

\newpage

{\bf{The $n=4,\ r=0\,$ case}}

\smallskip

Our next step will be to recover the results obtained in \cite{AS07} for the conformal blocks of four Fibonacci fields, assuming to this end $r = 0\,.$ In this case the limit (\ref{PrMr}) reads
\ba
&&\lim\limits_{z_i \to w_i}\, \prod_{1\le i < j \le 4} z^{- \frac{2}{3}}_{ij}
\prod_{1\le i\ne j \le 4} (w_i -z_j)^{- \frac{1}{3}}
\prod_{1\le i < j \le 4} w^{- \frac{1}{6}}_{ij}
%%= \nn\\
%%&&= \left( w_{12} w_{13} w_{14} w_{21} w_{23} w_{24} w_{31} w_{32} w_{34}
%%w_{41} w_{42} w_{43}\right)^{- \frac{1}{3}}\,\prod_{1\le i < j \le 4} w^{-    %%\frac{5}{6}}_{ij}   = \nn\\
= \left( w_{12} w_{34}\, w_{13} w_{24} \, w_{14} w_{23} \right)^{- \frac{3}{2}} = \nn\\
&&= %\mp
\left( w_{12} w_{34}\right)^{- \frac{9}{2}} (1-x)^{- \frac{3}{2}}\, x^3 =
%\pm
\prod_{i=1}^4 (w_i-z_i)^{-\frac{1}{3}}\left( w_{12} w_{34}\right)^{- \frac{9}{2}} (1-\eta)^{- \frac{3}{2}}\, \eta^3\ ,\qquad
\lb{Pr00}
\ea
where we have used (\ref{x}) and (\ref{eta-x}) to obtain the last two equalities.       
%\footnote{This is a good illustration of the fact that we have to be careful when %dealing with multivalued functions, e.g. assuming that $[(-1) (-1)]^{-\frac{3}{2}} = %1^{-\frac{3}{2}}\, \stackrel{?}{=} 1\,$ could lead to contradictions; in fact,
%$1^{-\frac{3}{2}} := e^{2k\pi i\, (-\frac{3}{2})} = e^{-3k\pi i} = (-1)^k\,, \
%k\in \N\,.$}.
Further, the terms in (\ref{PSI}) that survive in this limit are displayed in Table~\ref{tab1} (the last column of which will not be needed immediately).

\begin{table}[tbp]
\centering
\begin{tabular}{| l | l | l || l || l || l |} \hline
  $S_1$     & $S_2$ & $S_3$ & $\lim\limits_{z_i \to w_i}\,\Psi_{12,34}$ & $\lim\limits_{z_i \to w_i}\,\Psi_{13,24}$ &  $\lim\limits_{z_i \to w_i}\,\Psi_{14,23}$\\ \hline\hline
  $1 \,, 2$ & $3  $ & $4  $ & $0           $ & $0           $ & $w_{12}^2 (w_{13}w_{43}) (w_{24}w_{34})           $\\ \hline
  $1 \,, 2$ & $4  $ & $3  $ & $0           $ & $w_{12}^2 (w_{41}w_{43}) (w_{32}w_{34}) $& $0           $ \\ \hline
  $1 \,, 3$ & $2  $ & $4  $ & $0           $ & $0           $ & $w_{13}^2 (w_{12}w_{42}) (w_{24}w_{34})           $\\ \hline
  $1 \,, 3$ & $4  $ & $2  $ & $w_{13}^2 (w_{41}w_{42}) (w_{23}w_{24}) $ & $0                 $ & $0           $\\ \hline
  $1 \,, 4$ & $2  $ & $3  $ & $0           $ & $w_{14}^2 (w_{21}w_{23}) (w_{32}w_{34})          $ & $0           $\\ \hline
  $1 \,, 4$ & $3  $ & $2  $ & $w_{14}^2 (w_{31}w_{32}) (w_{23}w_{24}) $ & $0           $ & $0           $\\ \hline
  $2 \,, 3$ & $1  $ & $4  $ & $0           $ & $0           $ & $0           $\\ \hline
  $2 \,, 3$ & $4  $ & $1  $ & $w_{23}^2 (w_{41}w_{42}) (w_{13}w_{14}) $ & $w_{23}^2 (w_{41}w_{43}) (w_{12}w_{14})  $ & $0           $\\ \hline
  $2 \,, 4$ & $1  $ & $3  $ & $0           $ & $0 $ & $0           $\\ \hline
  $2 \,, 4$ & $3  $ & $1  $ & $w_{24}^2 (w_{31}w_{32}) (w_{13}w_{14}) $ & $0           $ & $w_{24}^2 (w_{13}w_{43}) (w_{21}w_{31})            $\\ \hline
  $3 \,, 4$ & $1  $ & $2  $ & $0           $ & $0 $ & $0           $\\ \hline
  $3 \,, 4$ & $2  $ & $1  $ & $0           $ & $w_{34}^2 (w_{21}w_{23}) (w_{12}w_{14})          $& $w_{34}^2 (w_{12}w_{42}) (w_{21}w_{41})           $ \\ \hline
\end{tabular}
\caption{\label{tab1} {The terms in $\Psi_{12,34}\,, \Psi_{13,24}\,$ and $\Psi_{14,23}\,$ for $r=0\,$ surviving in the limit $z_i \to w_i \,,\ i=1, \dots ,4 \ $}}
\end{table}

Putting everything together, we obtain
\ba
&&\lim\limits_{z_i\to w_i}\,\Psi_{12,34} (\{ w \} ; \{ z \}) = - \frac{2}{9}\, ( w_{13} w_{24} )  ( w_{14} w_{32} ) \left( w_{14} w_{32} + w_{13} w_{42} \right) = \nn\\
&&= -  \frac{2}{9}\, (w_{12} w_{34} )^3 \, \frac{(1-x)(2-x)}{x^3} =
\frac{2}{9}\, (w_{12} w_{34} )^3 \, \frac{(1-\eta)(2-\eta)}{\eta^3}
\ ,\nn\\
&&\lim\limits_{z_i\to w_i}\,\Psi_{13,24} (\{ w \} ; \{ z \}) = \frac{2}{9}\, (w_{12} w_{34}) (w_{14} w_{32}) \left( w_{14} w_{32} + w_{12} w_{34}\right) = \nn\\
&&= \frac{2}{9}\, (w_{12} w_{34} )^3 \, \frac{1+x}{x^2} =
\frac{2}{9}\, (w_{12} w_{34} )^3 \, \frac{(1-\eta)(1-2\eta)}{\eta^2}\ .
\lb{limPSI}\\
&&\lim\limits_{z_i\to w_i}\,\Psi_{14,23} (\{ w \} ; \{ z \}) = \frac{2}{9}\, (w_{12} w_{34}) (w_{13} w_{42}) \left( w_{12} w_{34} - w_{13} w_{42}\right) = \nn\\
&&= - \frac{2}{9}\, (w_{12} w_{34} )^3 \,\frac{(1-x)(1-2 x)}{x^2} =
- \frac{2}{9}\, (w_{12} w_{34} )^3 \,\frac{1+\eta}{\eta^2} \ .
\nn
\ea
It is easy to verify that the last expression for $\lim\limits_{z_i\to w_i}\,\Psi_{14,23}\,$ satisfies the linear relation (\ref{NWx}) whose counterpart in terms of $\eta\,$ (cf. (\ref{eta-x})) reads
\be
(1-\eta)\,\Psi_{14,23} (\{ w \} ; \{ z \}) = - \eta \, \Psi_{12,34} (\{ w \} ; \{ z \}) +  \Psi_{13,24} (\{ w \} ; \{ z \})\ .
\lb{NWeta}
\ee
We now have all the ingredients to perform the calculation of the correlator of four Fibonacci fields.

Formulae (A.18), (A.12) of \cite{AS07} for the two four point chiral conformal blocks of the Fibonacci field $\varepsilon (w)\,,\ \Delta_{\e} = \frac{2}{5}\,,$ expressed in terms of hypergeometric functions read
\ba
&&\Phi^{(0)}(\{w\}) := \langle\, \e (w_1) \dots \e (w_4) \, \rangle^{(0)} = \frac{1}{2} \left( w_{12} w_{34}\right)^{- \frac{4}{5}} (1-x)^{- \frac{2}{5}}\,\times \nn\\
&&\times \,[\, (2-x)\, F(\frac{1}{5} , -\frac{1}{5} , \frac{3}{5} ; x )\, + \, \frac{1}{3}\, x (1+x)\,F(\frac{6}{5} , \frac{4}{5} , \frac{8}{5} ; x ) \, ] \ ,\nn\\
&&\Phi^{(1)}(\{w\}) := \langle\, \e (w_1) \dots \e (w_4) \, \rangle^{(1)} = \frac{q^2\, C }{2} \,\left( w_{12} w_{34}\right)^{- \frac{4}{5}} x^{\frac{2}{5}} (1-x)^{- \frac{2}{5}}\,\times \nn\\
&& \times \,[\,(2-x)\, F(\frac{1}{5} , \frac{3}{5} , \frac{7}{5} ; x ) - 2  \, (1+x)\, F(\frac{1}{5} , \frac{3}{5} , \frac{2}{5} ; x )\, ]\ ,\lb{4e-1}\\
&&q := e^{i\frac{\pi}{5}}\ ,
\qquad 2\, C = \sqrt{\frac{\Gamma(\frac{1}{5}) \, \Gamma^3 (\frac{3}{5})}{ \Gamma (\frac{4}{5}) \, \Gamma^3 (\frac{2}{5})}}\ .\nn
\ea
It is straightforward to verify that they are reproduced by our own detailed calculations by choosing appropriately some sign factors (also needed to match the natural conditions (\ref{OPE40}) and (\ref{OPE41}), see below).
%\footnote{The normalizations of the two conformal blocks are independent anyway.}

Using the analytic continuation formula\footnote{All the information about hypergeometric functions needed in what follows can be found in any decent handbook on the subject like e.g. \cite{BE53} or \cite{AS72}.}
\be
F(a, b , c; \frac{\eta}{\eta -1} ) = (1-\eta)^a\, F(a, c - b , c; \eta)\ ,
\lb{HG-ac1}
\ee
we obtain from (\ref{AB1}) and (\ref{HGF})
\ba
&&A^{(0)} (\{ w \}) = (w_{12} w_{34})^{\frac{7}{10}}\, (1-\eta)^{\frac{1}{10}}\, F(\frac{1}{5}, \frac{4}{5}, \frac{3}{5}; \eta)\ ,\nn\\
&&B^{(0)} (\{ w \}) = - \frac{1}{3} \,(w_{12} w_{34})^{\frac{7}{10}}\, (1-\eta)^{\frac{1}{10}}\, F(\frac{6}{5}, \frac{4}{5}, \frac{8}{5}; \eta)\ , \nn\\
&&A^{(1)} (\{ w \}) = -\, C\, (w_{12} w_{34})^{\frac{7}{10}}\, \eta^{\frac{2}{5}}\, (1-\eta)^{-\frac{3}{10}}\, F(\frac{1}{5}, \frac{4}{5}, \frac{7}{5}; \eta) \ ,\nn\\
&&B^{(1)} (\{ w \}) = 2\, C \, (w_{12} w_{34})^{\frac{7}{10}}\, \eta^{-\frac{3}{5}}\, (1-\eta)^{-\frac{3}{10}}\, F(\frac{1}{5}, -\frac{1}{5}, \frac{2}{5}; \eta)\ ,
\lb{AB2}
\ea
and from (\ref{4e-1}),
\ba
&&\Phi^{(0)}(\{w\}) = \frac{1}{2} \left( w_{12} w_{34}\right)^{- \frac{4}{5}} (1-\eta)^{-\frac{2}{5}}\,\times\nn\\
&&\times[\, (2-\eta) F(\frac{1}{5} , \frac{4}{5} , \frac{3}{5} ; \eta )-
\frac{1}{3}\,\eta\, (1-2\eta)\,F(\frac{6}{5} , \frac{4}{5} , \frac{8}{5} ; \eta )\, ] \ ,\quad\lb{4e-20}\\
&&\Phi^{(1)}(\{w\})  = \frac{C }{2} \,\left( w_{12} w_{34}\right)^{- \frac{4}{5}} \eta^{\frac{2}{5}} (1-\eta)^{- \frac{4}{5}}\,\times\nn\\
&&\times\, [\,(2-\eta)\, F(\frac{1}{5} , \frac{4}{5} , \frac{7}{5} ; \eta ) -
2\, (1-2\eta)\,  F(\frac{1}{5} , -\frac{1}{5} , \frac{2}{5} ; \eta )\, ]\ .\qquad\quad
\lb{4e-21}
\ea
Formulae (\ref{4e-20}), (\ref{4e-21}) will be the starting point of our calculations that follow.

\medskip

The two conformal blocks $\Phi^{(p)}(\{w\}) \,,\ p=0,1\,$ are characterized by the following two basic properties (which fix their normalization as well).
\begin{itemize}
\item
The short distance asymptotics of the $\Phi\,$ basis vectors for $w_{12} \to 0\,$ reproduces the two- and the three-point function, respectively,
\ba
&&\Phi^{(0)}(\{w\}) = \langle\, \e (w_1) \dots \e (w_4) \rangle^{(0)}  \, \underset{w_{12}\to 0}{\sim}\,  w_{12}^{-\frac{4}{5}}\, \langle \,\e (w_3)\, \e (w_4) \rangle = ( w_{12} w_{34} )^{-\frac{4}{5}} \ ,\qquad\quad \lb{OPE40}\\
&&\Phi^{(1)}(\{w\}) = \langle \, \e (w_1) \dots \e (w_4) \rangle^{(1)}  \, \underset{w_{12}\to 0}{\sim}\, C_{\varepsilon' \varepsilon \varepsilon} \, w_{12}^{\frac{3}{5}}\,\langle\, \varepsilon' (w_2) \,\e (w_3)\, \e (w_4) \rangle = \qquad\qquad\,
\nn\\
&&= ( C_{\varepsilon' \varepsilon \varepsilon} )^2 \, (w_{12}w_{34})^{\frac{3}{5}}\, (w_{23} w_{24})^{-\frac{7}{5}}  \, \underset{w_{12}\to 0}{\sim}\,( C_{\varepsilon' \varepsilon \varepsilon} )^2 \, (w_{12}w_{34})^{- \frac{4}{5}}\,\eta^{\frac{7}{5}}\ ,\qquad\qquad\qquad\qquad\
\lb{OPE41}
\ea
in accord with the operator product expansion (OPE)
\be
\varepsilon (w_1) \,\varepsilon (w_2)\, \underset{w_{12}\to 0}{\sim}\,  w_{12}^{-\frac{4}{5}} \id + C_{\varepsilon' \varepsilon \varepsilon}  \, w_{12}^{\frac{3}{5}}\,  \varepsilon' (w_2)\ ,\qquad \Delta_{\e'} = \frac{7}{5}\ .
%\qquad {\rm or}\nn\\&&\e\,\e \sim \id + \e'\ ,
\lb{eeOPE}
\ee
%for short.
Similar conditions appear for $w_{34} \to 0\,.$

\medskip

\noindent
{\bf Remark 1~} One can infer from the OPE $\e'\,\e \sim Y + \e\,,\ \Delta_Y = 3\,$
(see Eq.(54) in \cite{D84} or Table 10.2 in \cite{DiFMS97})
that all $3$-point structure constants $C_{\varepsilon' \varepsilon \varepsilon}\,,\ C_{\varepsilon \varepsilon' \varepsilon}\,$ and $C_{\varepsilon \varepsilon \varepsilon'}\,$ are equal. Here is the complete list of non-trivial fusion relations in this sector of ($\Z_3\,$ Potts {\em thermal}) Virasoro fields:
\be
\e\,\e \sim \id + \e'\ ,\quad \e'\,\e \sim Y + \e\ ,\quad \e\, Y \sim \e'\ ,\quad \e'\, Y \sim \e\ ,\quad \e' \e' \sim \id + \e'\ ,\quad
Y\, Y \sim \id\ .
\lb{Yee}
\ee

\medskip

\noindent
{\bf Remark 2~} The short distance behavior (\ref{OPE40}) and (\ref{OPE41}) which we are going to prove below provides an "internal" characterization of the two channels, or conformal blocks indexed by $p = 0, 1,\,$ respectively. The fact that the channels, originally introduced with reference to the fusion (\ref{ss}) also correspond to the two possible outcomes in the OPE of two $\e\,$ fields (\ref{eeOPE}) will be used later to introduce the conformal blocks corresponding to a higher number of Fibonacci anyons, $n>4\,.$ (Of course, the two definitions are consistent.)

\medskip

\item The following two braidings (homotopy classes of analytic continuation) are diagonal in the $\Phi\,$ basis:
\ba
&&b_1 \equiv b_{12}: \ w_{12} \stackrel{\curvearrowleft}{\longrightarrow} w_{21} := e^{i\pi} w_{12}\ ,\quad
b_3 \equiv b_{34} : \ w_{34} \stackrel{\curvearrowleft}{\longrightarrow} w_{43} := e^{i\pi} w_{34}\nn\\
&&(b_i - q^{-4})\, \Phi^{(0)}(\{w\}) = 0\ ,\quad
(b_i - q^3)\, \Phi^{(1)}(\{w\}) = 0\ ,\ i = 1,3\ ,
\lb{B1B3}
\ea
or
\be
b_i \, \Phi(\{w\}) = R\, \Phi(\{w\})\ ,\quad
\Phi(\{w\}) :=  \begin{pmatrix} \Phi^{(0)}(\{w\})\cr \Phi^{(1)}(\{w\}) \end{pmatrix}\ , \quad
R = \begin{pmatrix} q^{-4} & 0 \cr 0 & q^3 \end{pmatrix}\ .
\lb{BBR}
\ee
\end{itemize}
The short distance asymptotics (\ref{OPE40}) and (\ref{OPE41}) are easily verified by taking into account the fact that $\eta \to 0\,$ if either $w_{12}\,$ or $w_{34}\,$ goes to zero and that the hypergeometric series expansion for small $\eta\,$ starts with
\be
F(a,b,c;\eta) = 1 + \frac{ab}{c}\, \eta + {\cal O} (\eta^2) \ .
\ee
The specific combination appearing in the expression (\ref{4e-21}) for
$\Phi^{(1)}(\{w\})\,$
\ba
&&(2-\eta )\, F(\frac{1}{5} , \frac{4}{5} , \frac{7}{5} ; \eta) -
2 (1-2\, \eta)\, F(\frac{1}{5} , -\frac{1}{5} , \frac{2}{5} ; \eta ) =
\lb{exp1}\\
&&= (2 - \eta )(1 + \frac{4}{35} \eta ) - 2 (1-2\, \eta ) (1 - \frac{1}{10}
 \eta ) + \dots = \frac{24}{7}\, \eta + {\cal O} (\eta^2)
\nn
\ea
provides, in particular, the additional (to $\eta^{\frac{2}{5}}$) power of $\eta\,$ needed to satisfy the last equality in (\ref{OPE41}), and also fixes the three-point structure constant
\be
C_{\varepsilon' \varepsilon \varepsilon}  = \sqrt{\frac{12}{7}\, C}\ .
\lb{Ceee}
\ee
As noted in \cite{AS07}, this detail is actually the manifestation of the field $\e' (w)\,$ of conformal dimension $\Delta_{\e'} = \frac{7}{5}\,$ whose appearance in the operator algebra of $\e (w)\,$ has been discovered long ago, see e.g. \cite{D84}.

It is obvious that the effect of the $b_1\,$ and $b_3\,$ braidings on the  $\Phi \,$ basis (\ref{4e-20}), (\ref{4e-21}) is the same since in both cases
\ba
&&(w_{12} w_{34})^{-\frac{4}{5}} \to (e^{i\pi} w_{12} w_{34})^{-\frac{4}{5}} = q^{-4} (w_{12} w_{34})^{-\frac{4}{5}}\ ,\qquad {\rm and}\nn\\
&&\eta =  \frac{w_{12} w_{34}}{w_{13} w_{24}} \to
- \frac{w_{12} w_{34}}{w_{23} w_{14}} = \frac{\eta}{\eta -1} \ \
(\, = \, e^{i\pi} \eta \, (1-\eta)^{-1}\, )\quad\Rightarrow\nn\\
&&1- \eta \to \frac{1}{1-\eta}\ ,\quad 2-\eta \to \frac{2 -\eta}{1-\eta}\ ,\quad 1 - 2\,\eta \to \frac{1+\eta}{1-\eta}\ .
\lb{B1B3Phi}
\ea
So, for example, we obtain from (\ref{4e-20}) by using (\ref{HG-ac1}) and (\ref{B1B3Phi})
\ba
&&b_1 \Phi^{(0)}(\{w\}) = \frac{q^{-4}}{2} \left( w_{12} w_{34}\right)^{- \frac{4}{5}} (1-\eta)^{\frac{2}{5}} \times\nn\\
&&\times [\, \frac{2-\eta}{1-\eta} (1-\eta)^{\frac{1}{5}} F(\frac{1}{5} , -\frac{1}{5} , \frac{3}{5} ; \eta ) + \frac{1}{3}\frac{\eta\, (1+\eta)}{(1-\eta)^2}\,(1-\eta)^{\frac{6}{5}} F(\frac{6}{5} , \frac{4}{5} , \frac{8}{5} ; \eta ) \, ] \ ,\qquad\qquad \lb{B10}
\ea
and it remains to apply one of the Gauss' contiguous relations,
\be
F(\frac{1}{5} , -\frac{1}{5} , \frac{3}{5} ; \eta ) =
F(\frac{1}{5} , \frac{4}{5} , \frac{3}{5} ; \eta ) - \frac{1}{3}\,\eta\,
F(\frac{6}{5} , \frac{4}{5} , \frac{8}{5} ; \eta )
\lb{Gcr1}
\ee
to confirm the first equality in (\ref{B1B3}), the one for the braiding
\begin{equation*}
b_1 \Phi^{(0)}(\{w\}) = q^{-4}\,\Phi^{(0)}(\{w\})\ .
\end{equation*}
Similarly,
\ba
&&b_1 \Phi^{(1)}(\{w\})  = \frac{C }{2} \,q^{-4} (w_{12} w_{34})^{- \frac{4}{5}} q^2\, \eta^{\frac{2}{5}} (1-\eta)^{\frac{2}{5}} \times\nn\\
&&\times\,[\, \frac{2-\eta}{1-\eta}\, (1-\eta)^{\frac{1}{5}}  F(\frac{1}{5} , \frac{3}{5} , \frac{7}{5} ; \eta ) - 2\, \frac{1+\eta}{1-\eta}\,(1-\eta)^{\frac{1}{5}}  F(\frac{1}{5} , \frac{3}{5} , \frac{2}{5} ; \eta )\, ]\ =\nn\\
&&= q^{-2}\,\frac{C }{2}(w_{12} w_{34})^{- \frac{4}{5}}  \eta^{\frac{2}{5}} (1-\eta)^{-\frac{2}{5}} \times \lb{B11}\\
&&\times\,[\, (2-\eta) \, (1-\eta)^{\frac{3}{5}}  F(\frac{6}{5} , \frac{4}{5} , \frac{7}{5} ; \eta ) - 2\, (1+\eta)\,(1-\eta)^{-\frac{2}{5}}  F(\frac{1}{5} , - \frac{1}{5} , \frac{2}{5} ; \eta )\, ]\ ,\nn
\ea
the last expression following from the previous one due to
\be
F (a,b,c ; \eta) = (1-\eta)^{c-a-b}\,F(c-a, c-b, c ; \eta)\ .
\lb{HGrel}
\ee
In the final step we apply another Gauss' contiguous relation,
\be
(1-\eta)\,F(\frac{6}{5} , \frac{4}{5} , \frac{7}{5} ; \eta ) = -
F(\frac{1}{5} , \frac{4}{5} , \frac{7}{5} ; \eta ) + 2\,
F(\frac{1}{5} , - \frac{1}{5} , \frac{2}{5} ; \eta )
\lb{Gcr3}
\ee
to obtain also the second equality in (\ref{B1B3}),
\begin{equation*}
b_1 \Phi^{(1)}(\{w\}) = - q^{-2}\,\Phi^{(1)}(\{w\}) = q^3\,\Phi^{(1)}(\{w\})\ .
\end{equation*}

\section{The dual basis and the fusion matrix ($n=4\,,\ r=0$)}

\setcounter{equation}{0}
\renewcommand\theequation{\thesection.\arabic{equation}}

The $\Phi\,$ basis (\ref{4e-20}), (\ref{4e-21}) of four-point Fibonacci field conformal blocks is well adapted to study the $\eta \sim 0\,$ behaviour (which means small $w_{12}\,$ or $w_{34}$). The braiding of the two middle fields is however related to small $w_{23}\,$ or, equivalently, $1-\eta \sim 0\,,$ i.e. $\eta \sim 1\,.$ This requires the introduction of a "dual" basis (denoted as $\T\,$ below) the vectors of which correspond to the two channels appearing after fusing the second and third $\e\,$ fields (and not the first and second, or the third and the fourth one, as it is assumed in the construction of the $\Phi\,$ basis)\footnote{This situation is very well known in high energy physics where the "dual" description of four-point scattering amplitudes (the $\Phi\,$ and $\T\,$ bases corresponding to the $s$- and $u$-channels, respectively, following the standard notation for the Mandelstam variables) led to the Veneziano formula (1968) and, subsequently, to the first idea of using string theory in the form of the so called dual resonance model of strong interactions.}.

Thus, technically the dual basis $\T(\{w\}) = \begin{pmatrix} \T^{(0)}(\{w\}) \cr \T^{(1)}(\{w\}) \end{pmatrix} \,$ has to be determined by the following three conditions:
\begin{itemize}
\item
The vectors $\T^{(p)}(\{w\})\,,\ p = 0,1\,$ are linear combinations of $\Phi^{(q)}(\{w\})\,,\ q = 0,1\,.$
\item
The short distance asymptotics of the $\T\,$ basis vectors for $w_{23} \to 0\,$ reproduces the corresponding two- and the three-point function (compare with (\ref{OPE40}) and (\ref{OPE41}) and note that $1-\eta = \frac{w_{23}w_{14}}{w_{13} w_{24}}\,$)\,:
\ba
%= \langle\, \e (w_1) \dots \e (w_4) \rangle_{(0)}  \, \underset{w_{23}}
&&\T^{(0)}(\{w\})  \, \underset{w_{23}\to 0}{\sim}\,  w_{23}^{-\frac{4}{5}}\, \langle \,\e (w_1) \, \e (w_4) \rangle = ( w_{23} w_{14} )^{-\frac{4}{5}} \ ,\,\lb{OPE40p}\\
&&\T^{(1)}(\{w\})
%= \langle \, \e (w_1) \dots \e (w_4) \rangle^{(1)}  \,
\underset{w_{23}\to 0}{\sim}\, C_{\varepsilon \varepsilon' \varepsilon} \, w_{23}^{\frac{3}{5}}\,\langle\, \e (w_1)\, \e' (w_3)\, \e (w_4) \rangle = \qquad\qquad\,\ \nn\\
&&= ( C_{\varepsilon \varepsilon' \varepsilon} )^2 \, (w_{23}w_{14})^{\frac{3}{5}}\, (w_{13} w_{34})^{-\frac{7}{5}}  \, \underset{w_{23}\to 0}{\sim}\, \frac{12}{7} \, C\, (w_{23}w_{14})^{- \frac{4}{5}}\,(1-\eta)^{\frac{7}{5}}\ .
\qquad
\lb{OPE41p}
\ea
\item The braiding
\be
b_2 \equiv b_{23}: \ w_{23} \stackrel{\curvearrowleft}{\longrightarrow} w_{32} := e^{i\pi} w_{23}
\lb{B2s}
\ee
is diagonal in the (dual) $\T\,$ basis with the same eigenvalues as in (\ref{BBR})\,:
\be
b_2 \, \T(\{w\}) = R\, \T(\{w\})\, ,\quad
R = \begin{pmatrix} q^{-4} & 0 \cr 0 & q^3 \end{pmatrix}\ .
\lb{BBRs}
\ee
\end{itemize}

To this end we will start by performing an innocent procedure by just recasting the expressions for $\Phi^{(q)}(\{w\})\,,\ q = 0,1\,$ (\ref{4e-20}), (\ref{4e-21}) replacing $ w_{12} w_{34}\,$ with $w_{23} w_{14}\,,$
\begin{equation*}
w_{12} w_{34} = w_{23} w_{14} \, \frac{\eta}{1-\eta}
\end{equation*}
as well the argument $\eta\,$ of hypergeometric series with $1-\eta\,$ by using the equality
\ba
&&F(a,b,c;\eta) = \frac{\G (c)\,\G (c-a-b)}{\G (c-a)\,\G (c-b)}
\,F(a,b,a+b-c+1; 1-\eta) +  \lb{HGFancont1}\\
&&+ \,\frac{\G (c)\,\G (a+b-c)}{\G (a)\,\G (b)}\,(1-\eta)^{c-a-b}\, F(c-a,c-b,c-a-b+1; 1-\eta)
\nn
\ea
which gives
\ba
F(\frac{1}{5} , \frac{4}{5} , \frac{3}{5} ; \eta ) &=& \eta^{\frac{2}{5}}\,[\tau \, (1-\eta)^{-\frac{2}{5}}\,F(\frac{1}{5} , \frac{4}{5} , \frac{3}{5} ; 1-\eta ) + \frac{D_1}{2}\,F(\frac{6}{5} , \frac{3}{5} , \frac{7}{5} ; 1-\eta)\,]\ ,\nn\\
F(\frac{6}{5} , \frac{4}{5} , \frac{8}{5} ; \eta ) &=& 3\,\eta^{-\frac{3}{5}}\,[\,- \frac{D_1}{2}\,F(\frac{1}{5} , \frac{3}{5} , \frac{7}{5} ; 1-\eta) + \tau \, (1-\eta)^{-\frac{2}{5}}\,F(\frac{1}{5} , - \frac{1}{5} , \frac{3}{5} ; 1-\eta)\,]\ ,
\nn\\
F(\frac{1}{5} , \frac{4}{5} , \frac{7}{5} ; \eta ) &=& - \tau \,(1-\eta)^{\frac{2}{5}}\,F(\frac{6}{5} , \frac{3}{5} , \frac{7}{5} ; 1-\eta)
+ 2 \,D_2\,F(\frac{1}{5} , \frac{4}{5} , \frac{3}{5} ; 1-\eta )  \ ,
\lb{HGac3}\\
F(\frac{1}{5} , -\frac{1}{5} , \frac{2}{5} ; \eta ) &=&
\frac{1}{2}\,
[\,2\,D_2 \, F(\frac{1}{5} , - \frac{1}{5} , \frac{3}{5} ; 1-\eta) +
\tau \,(1-\eta)^{\frac{2}{5}} \,
F(\frac{1}{5} , \frac{3}{5} , \frac{7}{5} ; 1-\eta )  \, ] \ .
\nn
\ea
We have applied twice a version of (\ref{HGrel}) above to replace
\ba
&&F(\frac{2}{5} , -\frac{1}{5} , \frac{3}{5} ; 1-\eta)\qquad {\rm by}\qquad \eta^{\frac{2}{5}}\, F(\frac{1}{5} , \frac{4}{5} , \frac{3}{5} ; 1-\eta)\ ,\qquad{\rm and}\nn\\
&&F(\frac{1}{5} , \frac{4}{5} , \frac{7}{5} ; 1-\eta)\qquad\ \ \, {\rm by}\qquad \eta^{\frac{2}{5}}\, F(\frac{6}{5} , \frac{3}{5} , \frac{7}{5} ; 1-\eta)\ .\nn
\ea
In (\ref{HGac3}) $\tau = \frac{\sqrt{5}-1}{2}\,$ is the inverse of the golden ratio, or
\ba
&&\tau =
\frac{\G (\frac{2}{5})\G (\frac{3}{5})}{\G (\frac{1}{5})\,\G (\frac{4}{5})}=
\frac{1}{2 \,\cos{\frac{\pi}{5}}} = \frac{1}{q+q^{-1}} = q^2 + q^{-2} \quad ( \tau^2 + \tau = 1 )\ ,\quad {\rm and}
\lb{tau}
\nn\\
&&D_1 := \frac{\G^2 (\frac{3}{5})}{\G (\frac{2}{5})\, \G (\frac{4}{5})} = 2\,C \,\sqrt{\tau}\ ,\qquad
D_2 := \frac{\G^2 (\frac{2}{5})}{ \G (\frac{1}{5})\,  \G (\frac{3}{5}) } = \frac{1}{2\,C}\,\sqrt{\tau} \ ,\qquad
\lb{D2}
\ea
with $C\,$ as given in (\ref{4e-1}). Putting everything together, we obtain the desired presentation of the $\Phi\,$ basis in the form
\ba
&&\Phi^{(0)} (\{w\} ) = \frac{1}{2} \left( w_{23} w_{14}\right)^{- \frac{4}{5}} \eta^{-\frac{2}{5}} \times\nn\\
&&\times \{\, \tau\, [\, (2-\eta)\, F(\frac{1}{5} , \frac{4}{5} , \frac{3}{5} ; 1-\eta )- (1-2\eta)\,F(\frac{1}{5} , -\frac{1}{5} , \frac{3}{5} ; 1-\eta )\, ] +\lb{Phi0-F}\\
&&+\, C\,\sqrt{\tau}\, (1-\eta)^{\frac{2}{5}}\,[\, (2-\eta)\, F(\frac{6}{5} , \frac{3}{5} , \frac{7}{5} ; 1-\eta ) + (1-2\eta)\,F(\frac{1}{5} , \frac{3}{5} , \frac{7}{5} ; 1-\eta )\, ] \,\}\ ,\nn\\
&&\Phi^{(1)} (\{ w\} ) = \frac{1}{2} \left( w_{23} w_{14}\right)^{- \frac{4}{5}} \eta^{-\frac{2}{5}} \times\nn\\
&&\times \{\, \sqrt{\tau}\, [\, (2-\eta)\, F(\frac{1}{5} , \frac{4}{5} , \frac{3}{5} ; 1-\eta )- (1-2\eta)\,F(\frac{1}{5} , -\frac{1}{5} , \frac{3}{5} ; 1-\eta )\, ] - \nn\\
\lb{Phi1-F}
&& - \,C\,\tau\, [\, (2-\eta)\, F(\frac{6}{5} , \frac{3}{5} , \frac{7}{5} ; 1-\eta ) + (1-2\eta)\,F(\frac{1}{5} , \frac{3}{5} , \frac{7}{5} ; 1-\eta )\, ] \,\}\ .
\ea
A careful look reveals that our simple exercise actually produced an amazing result, since (\ref{Phi0-F}), (\ref{Phi1-F}) can be written compactly as
\ba
&&\begin{pmatrix} \Phi^{(0)}(\{w\}) \cr \Phi^{(1)}(\{w\}) \end{pmatrix}
= \begin{pmatrix} \tau&\sqrt{\tau} \cr \sqrt{\tau}&- \tau\end{pmatrix}\  \begin{pmatrix} \T^{(0)}(\{w\}) \cr \T^{(1)}(\{w\}) \end{pmatrix}
\ ,\qquad{\rm or}\nn\\
&&\Phi (\{ w\} ) = F\, \T (\{ w\} )\ ,\qquad F = \begin{pmatrix} \tau&\sqrt{\tau} \cr \sqrt{\tau}&- \tau\end{pmatrix}
\lb{PhiFPsi}
\ea
(note the involutivity of the matrix $F\,,\ F^2 = \id\,$ for $\tau\,$ given by (\ref{tau})), with
\ba
&&\T^{(0)}(\{w\}) = \frac{1}{2} \left( w_{23} w_{14}\right)^{- \frac{4}{5}} \eta^{-\frac{2}{5}} \times\nn\\
&&\times \, [\, (2-\eta)\, F(\frac{1}{5} , \frac{4}{5} , \frac{3}{5} ; 1-\eta )- (1-2\eta)\,F(\frac{1}{5} , -\frac{1}{5} , \frac{3}{5} ; 1-\eta )\, ]\ ,\lb{Psi0-b2}\\
&&\T^{(1)}(\{w\}) = \frac{C}{2} \left( w_{23} w_{14}\right)^{- \frac{4}{5}} \eta^{-\frac{2}{5}} \,(1-\eta)^{\frac{2}{5}}  \times\nn\\
&&\times \, [\, (2-\eta)\, F(\frac{6}{5} , \frac{3}{5} , \frac{7}{5} ; 1-\eta ) + (1-2\eta)\,F(\frac{1}{5} , \frac{3}{5} , \frac{7}{5} ; 1-\eta )\, ]\ ,
\lb{Psi1-b2}
\ea
and the matrix $F\,$ coincides with the canonical solution of the relevant pentagon equation (see e.g. Eq.(9.125) in J. Preskill's
Lecture Notes \cite{P04}). This fact strongly suggests that $\T^{(0)}(\{w\})\,,\ \T^{(1)}(\{w\})\,$ (\ref{Psi0-b2}), (\ref{Psi1-b2}) form the dual basis we have been looking for.

We will now prove that $\T^{(0)}(\{w\})\,$ and $\T^{(1)}(\{w\})\,$ satisfy indeed the requirements spelled out in Eqs. (\ref{OPE40p}), (\ref{OPE41p}) and (\ref{BBRs}).

The verification of the $w_{23} \to 0\,$ (and hence, $\eta \to 1\,$) asymptotics goes quite similarly to the $\Phi\,$ basis case. After substituting
\begin{equation*}
1-\eta = \epsilon\quad\Rightarrow\quad 2-\eta = 1+ \epsilon\ ,\quad
1-2\,\eta = 2\,\epsilon - 1\qquad (\epsilon \to 0)\,,
\end{equation*}
it amounts to showing that
\be
2- \eta - (1 - 2\eta ) + {\cal O} (1-\eta) = 1+\epsilon - (2\,\epsilon -1) + {\cal O} (\epsilon) = 2 + {\cal O} (\epsilon)
\lb{OPE40u}
\ee
and
\ba
&&(2-\eta) (1 + \frac{18}{35} (1-\eta) ) + (1-2\, \eta)(1+ \frac{3}{35} (1-\eta)) + {\cal O} ((1-\eta)^2) = \nn\\
&&= (1+\epsilon) (1 + \frac{18}{35}\, \epsilon ) + (2\, \epsilon - 1)(1+ \frac{3}{35} \epsilon) + {\cal O} (\epsilon^2) = \frac{24}{7}\,\epsilon
+ {\cal O} (\epsilon^2)\ ,\qquad\quad
\lb{OPE41u}
\ea
respectively.

To prove that the action of the braiding $b_2\,$ (\ref{B2s}) on the $\T\,$ basis (\ref{Psi1-b2}), (\ref{Psi1-b2}) is given by the diagonal matrix $R\,,$ we proceed as follows. As the exchange of $w_2\,$ and $w_3\,$ induces
\be
\eta = \frac{w_{12}w_{34}}{w_{13}w_{24}}\to \frac{w_{13}w_{24}}{w_{12}w_{34}} = \eta^{-1}\quad\Rightarrow\quad 1-\eta\ \to \  1-\eta^{-1} = \frac{1-\eta}{(1-\eta) - 1}\ ,\quad
\lb{B2eta}
\ee
we need a version of the relation (\ref{HG-ac1}) in the form
\be
F(a, b , c; 1-\eta^{-1} ) = \eta^a\, F(a, c - b , c; 1-\eta)\ .
\lb{HG-ac11}
\ee
Applying it to (\ref{Psi0-b2}), we get
\ba
&&b_2\, \T^{(0)}(\{w\}) = \frac{1}{2}\, q^{-4} \,\left( w_{23} w_{14}\right)^{- \frac{4}{5}} \eta^{\frac{2}{5}} \times\nn\\
&&\times \, [\, - \frac{1-2\,\eta}{\eta}\, \eta^{\frac{1}{5}}\,F(\frac{1}{5} , -\frac{1}{5} , \frac{3}{5} ; 1-\eta ) + \frac{2-\eta}{\eta}\, \eta^{\frac{1}{5}}\,F(\frac{1}{5} , \frac{4}{5} , \frac{3}{5} ; 1-\eta )\, ]\, =\nn\\
&&= q^{-4}\,\T^{(0)}(\{w\})\ .
\lb{Psi0-b}
\ea
In the case of (\ref{Psi1-b2}) we obtain
\ba
&&b_2\, \T^{(1)}(\{w\}) = \frac{C}{2} \,q^{-4}\,\left( w_{23} w_{14}\right)^{- \frac{4}{5}} \eta^{\frac{2}{5}} \,q^2 \,\left(\frac{1-\eta}{\eta}\right)^{\frac{2}{5}}  \times\nn\\
&&\times \, [\, - \frac{1-2\,\eta}{\eta}\,\eta^{\frac{6}{5}}\,  F(\frac{6}{5} , \frac{4}{5} , \frac{7}{5} ; 1-\eta ) - \frac{2- \eta}{\eta}\,\eta^{\frac{1}{5}}\,F(\frac{1}{5} , \frac{4}{5} , \frac{7}{5} ; 1-\eta )\, ] = \nn\\
&&= - \, q^{-2} \, \frac{C}{2} \,\left( w_{23} w_{14}\right)^{- \frac{4}{5}}
\,\eta^{- \frac{4}{5}} \,(1-\eta)^{\frac{2}{5}}\times\nn\\
&&\times \, [\, (2- \eta) \, F(\frac{1}{5} , \frac{4}{5} , \frac{7}{5} ; 1-\eta ) + (1-2\,\eta)\,\eta\,  F(\frac{6}{5} , \frac{4}{5} , \frac{7}{5} ; 1-\eta )  \,] = \nn\\
&&= q^3 \, \T^{(1)}(\{w\}) \ ,
\lb{Psi1-b}
\ea
the last equality taking place due to (\ref{HGrel}) which gives
\be
F(\frac{1}{5} , \frac{4}{5} , \frac{7}{5} ; 1-\eta ) = \eta^{\frac{2}{5}}\, F(\frac{6}{5} , \frac{3}{5} , \frac{7}{5} ; 1-\eta )\ , \quad
F(\frac{6}{5} , \frac{4}{5} , \frac{7}{5} ; 1-\eta ) = \eta^{- \frac{3}{5}}\, F(\frac{1}{5} , \frac{3}{5} , \frac{7}{5} ; 1-\eta )\ .
\lb{HGac5}
\ee
We have thus confirmed (\ref{BBRs}) in the dual basis. Together with (\ref{BBR}),
(\ref{PhiFPsi}) (and $F = F^{-1}\,$) it implies that the three generators of the Artin braid group ${\cal B}_4\,$ acting on the $\Phi\,$ basis are represented by the following matrices:
\be
\pi^{(4)} (b_1) = \pi^{(4)} (b_3) = R = \begin{pmatrix} q^{-4} & 0 \cr 0 & q^3 \end{pmatrix}\ ,
\quad \pi^{(4)} (b_2) = B\ ,\quad B:= F\,R\,F = \begin{pmatrix} q^4 \tau & q^{-3} \sqrt{\tau} \cr q^{-3} \sqrt{\tau} & - \tau\end{pmatrix}\ .
\lb{B3Phi}
\ee
(Of course, in the dual, $\T\,$ basis in which $b_2\,$ acts diagonally by $R\,,\ b_1\,$ and $b_3\,$ are represented by $B\,.$) To make sure that the generators given by (\ref{B3Phi}) satisfy the Artin relations for ${\cal B}_4\,$
\be
b_1\, b_2\, b_1 = b_2\, b_1\, b_2\ ,\quad b_2\, b_3\, b_2 = b_3\, b_2\, b_3\ ,\quad b_1\, b_3 = b_3\, b_1\ ,
\lb{ArtinB4}
\ee
we need to verify the matrix equality
\be
R \, B\,  R = B \, R \, B \ \qquad ( {\rm for}\  B = F\, R\, F )\ .
\lb{Artin}
\ee
To show that (\ref{Artin}) holds, one can express $\tau\,$ in terms of $q\,$ using (\ref{tau}); it turns out indeed that both sides are equal (to $q^{-4}\, F$). Note also that
\be
\det F = - 1\ ,\qquad \det R = q^{-1} = \det B\ ,
\lb{dets}
\ee
the equality of the last two determinants being actually a consistency condition for (\ref{Artin}).

\section{Braiding and fusion for $n=4\,$ and arbitrary $r$}

\setcounter{equation}{0}
\renewcommand\theequation{\thesection.\arabic{equation}}

In the presence of $3\,r\,$ electrons at points $z_5 , \dots , z_{3 r + 4}\,$  the prefactor (\ref{PrMr}) (for $M=0$) contains the product
\begin{equation*}
\prod_{1\le i<j \le 4} w_{ij}^{-\frac{3}{2}} =
(w_{12} w_{34})^{-\frac{9}{2}}\,\eta^3\, (1-\eta)^{-\frac{3}{2}}\ ,
\end{equation*}
see (\ref{eta-x}) and, in addition (after the fusion limit is taken, cf. (\ref{OPEspe})), the piece
\be
Q = Q(\{w\}, \{\underline{z}\}) := \frac{9}{4} \,\prod_{i=1}^4 \prod_{j=5}^{3r+4} (w_i-z_j)^{-1} \,\prod_{5\le k < \ell \le 3r+4}z_{k\ell}^{- \frac{2}{3}}\ ,
\quad \{\underline{z}\}\,\equiv\, \{ z_5, \dots, z_{3r+4}\}
\lb{Qwuz}
\ee
which is symmetric in $w_i\,,\ i=1,2,3,4\,.$
Accordingly (fixing the sign of $\Phi^{(1)}(\{w\}, \{ \uz \})$),
\ba
&&\Phi^{(0)}(\{w\}, \{ \uz \} ) = Q \,[(w_{12} w_{34})^{-3}\,\eta^3\, (1-\eta)^{-\frac{3}{2}} ]\, \left( w_{12} w_{34}\right)^{- \frac{4}{5}} (1-\eta)^{\frac{1}{10}}\,\times\nn\\
&&\times\, [\, F(\frac{1}{5} , \frac{4}{5} , \frac{3}{5} ; \eta )\, \Psi_{12,34} -\frac{1}{3} \,F(\frac{6}{5} , \frac{4}{5} , \frac{8}{5} ; \eta )\, \Psi_{13,24}\, ] \ ,\qquad\ \lb{4e-eta0}\\
&&\Phi^{(1)}(\{w\}, \{ \uz \})  = C\,Q\,[(w_{12} w_{34})^{-3}\,\eta^3\, (1-\eta)^{-\frac{3}{2}} ]\left( w_{12} w_{34}\right)^{- \frac{4}{5}} \eta^{-\frac{3}{5}} (1-\eta)^{- \frac{3}{10}}\,\times\nn\\
&&\times\, [\,\eta\, F(\frac{1}{5} , \frac{4}{5} , \frac{7}{5} ; \eta )\,\Psi_{12,34} - 2 \,F(\frac{1}{5} , -\frac{1}{5} , \frac{2}{5} ; \eta )\, \Psi_{13,24}\, ]\ .
\lb{4e-eta1}
\ea
Note that $(w_{12} w_{34})^{-3}\,\eta^3\, (1-\eta)^{-\frac{3}{2}}\,$ is invariant with respect to $w_{12} \to e^{i\pi}\,w_{12}\,.$ Here
\begin{equation*}
\Psi_{12,34} = \Psi_{12,34}(\{w\}, \{w,\underline{z}\})\qquad {\rm and }\qquad
\Psi_{13,24}= \Psi_{13,24}(\{w\}, \{w,\underline{z}\})\ 
\end{equation*}
are the values at $z_i=w_i\,,\ i=1,2,3,4\,$ of the corresponding polynomials (\ref{PSI}) satisfying (\ref{NWeta}) so that
\be
\frac{\eta}{\eta-1}\, \, \Psi_{12,34} + \frac{1}{1-\eta}\,\Psi_{13,24} = \Psi_{14,23} \quad ( = \Psi_{14,23}(\{w\}, \{w,\underline{z}\})\,)\ .
\lb{NWeta-w}
\ee
(Obviously, exchanging the arguments in polynomials is path independent so that braiding reduces to permutation.) We will use in what follows (\ref{NWeta-w}) to find the braid group representation acting on the conformal blocks.

Of course, this alternative technique is also applicable to the special case $r=0\,$ when (\ref{4e-eta0}), (\ref{4e-eta1}) reduce, by (\ref{limPSI}) (and $Q = \frac{9}{4}$)\ to (\ref{4e-20}), (\ref{4e-21}).

\smallskip

{\bf N.B.} We emphasize that, in the presence of electrons (at 
points $\{ \uz \}$ (\ref{Qwuz})), the braiding only applies to the anyons with coordinates $\{ w_i \}_{i=1}^4\,.$  

\smallskip

For $w_{12} \to e^{i\pi} w_{12}\,$ (or $w_{34} \to e^{i\pi} w_{34}$) we have, by (\ref{HG-ac1}) and (\ref{NWeta-w}),% and (\ref{HGrel}),
\ba
&&\phi^{(0)} := F(\frac{1}{5} , \frac{4}{5} , \frac{3}{5} ; \eta )\, \Psi_{12,34} -\frac{1}{3} \,F(\frac{6}{5} , \frac{4}{5} , \frac{8}{5} ; \eta )\, \Psi_{13,24}\quad \to\quad (1-\eta)^{\frac{1}{5}} \phi^{(0)}\ ,\qquad \lb{B101}\\
&&\phi^{(1)} :=  \eta\, F(\frac{1}{5} , \frac{4}{5} , \frac{7}{5} ; \eta )\,\Psi_{12,34} - 2 \,F(\frac{1}{5} , -\frac{1}{5} , \frac{2}{5} ; \eta )\, \Psi_{13,24} \quad\to\quad (1-\eta)^{- \frac{6}{5}} \phi^{(1)}\qquad\quad
\lb{B111}
\ea
where we have used a Gauss' contiguous relation and (\ref{HGrel}) to derive
\ba
&&2\,F(\frac{1}{5} , \frac{3}{5} , \frac{2}{5} ; \eta ) -
F(\frac{1}{5} , \frac{3}{5} , \frac{7}{5} ; \eta ) =
F(\frac{6}{5} , \frac{3}{5} , \frac{7}{5} ; \eta ) = (1-\eta)^{-\frac{2}{5}}
F(\frac{1}{5} , \frac{4}{5} , \frac{7}{5} ; \eta )\ ,\nn\\
&&F(\frac{1}{5} , \frac{3}{5} , \frac{2}{5} ; \eta ) = (1-\eta)^{-\frac{2}{5}}
F(\frac{1}{5} , - \frac{1}{5} , \frac{2}{5} ; \eta )\ .
\lb{Gcr2}
\ea
This generalizes (\ref{BBR}) to arbitrary $r\,$ with the same matrix
$R = \begin{pmatrix} q^{-4} & 0 \cr 0 & q^3 \end{pmatrix}\,:$
\be
b_i \, \Phi(\{w\}, \{\uz\}) = R\, \Phi(\{w\}, \{\uz\})\,,\quad
\Phi(\{w\}, \{\uz\}) :=  \begin{pmatrix} \Phi^{(0)}(\{w\}, \{\uz\})\cr
\Phi^{(1)}( \{w\}, \{\uz\}) \end{pmatrix}\ , \ \quad i=1,3\ .
\lb{BBRgen}
\ee
To find the dual basis for arbitrary $r\,,$ we proceed as in the special case $r=0\,.$ We first recast (\ref{4e-eta0}), (\ref{4e-eta1}) by using (\ref{HGac3}):
\ba
&&\Phi^{(0)}(\{w\}, \{ \uz \} ) = Q \,[\,(w_{12} w_{34})^{-3}\,\eta^3\, ] \left( w_{23} w_{14}\right)^{- \frac{4}{5}} \eta^{-\frac{7}{5}} (1-\eta)^{-\frac{3}{5}}\,\times \nn\\
&&\times \{\, \eta\, [\, \tau\, (1-\eta)^{- \frac{2}{5}}\, F(\frac{1}{5} , \frac{4}{5} , \frac{3}{5} ; 1-\eta ) + \frac{D_1}{2}\,F(\frac{6}{5} , \frac{3}{5} , \frac{7}{5} ; 1-\eta )\, ] \, \Psi_{12,34} +\nn\\
&&+\,[\,\frac{D_1}{2}\, F(\frac{1}{5} , \frac{3}{5} , \frac{7}{5} ; 1-\eta ) - \tau\,(1-\eta)^{- \frac{2}{5}}\, F(\frac{1}{5} , - \frac{1}{5} , \frac{3}{5} ; 1-\eta )\, ]\, \Psi_{13,24}\,\}\ ,\qquad \lb{Phi0-Fgen}\\
&&\Phi^{(1)}(\{w\}, \{ \uz \} ) = C\,Q \,[\,(w_{12} w_{34})^{-3}\,\eta^3\, ] \left( w_{23} w_{14}\right)^{- \frac{4}{5}} \eta^{-\frac{7}{5}} (1-\eta)^{-1}\,\times \nn \\
&&\times \{\, \eta\, [\,- \sqrt{\tau}\,(1-\eta)^{\frac{2}{5}} F(\frac{6}{5} , \frac{3}{5} , \frac{7}{5} ; 1-\eta ) + 2\, D_2\,F(\frac{1}{5} ,\frac{4}{5} , \frac{3}{5} ; 1-\eta )\, ]\,\Psi_{12,34} - \nn\\
&& - [\,2\,D_2\, (1-\eta)^{\frac{2}{5}}\, F(\frac{1}{5} , - \frac{1}{5} , \frac{3}{5} ; 1-\eta ) + \tau \,(1-\eta)^{\frac{2}{5}} \,F(\frac{1}{5} , \frac{3}{5} , \frac{7}{5} ; 1-\eta )\, ]\,\Psi_{13,24}\,\}\ .\qquad\qquad
\lb{Phi1-Fgen}
\ea
This gives
\be
\begin{pmatrix} \Phi^{(0)}(\{w\}, \{ \uz \}) \cr \Phi^{(1)}(\{w\}, \{ \uz \}) \end{pmatrix}
= \begin{pmatrix} \tau&\sqrt{\tau} \cr \sqrt{\tau}&- \tau\end{pmatrix}\  \begin{pmatrix} \T^{(0)}(\{w\}, \{ \uz \}) \cr \T^{(1)}(\{w\}, \{ \uz \}) \end{pmatrix}
\lb{PhiFPsi-gen}
\ee
with
\ba
&&\T^{(0)}(\{w\}, \{ \uz \}) = Q\,[\,(w_{12} w_{34})^{-3}\,\eta^3\, ] \left( w_{23} w_{14}\right)^{- \frac{4}{5}} \eta^{-\frac{7}{5}} (1-\eta)^{-1}\,\times \nn\\
&&\times \, [\, \eta\,F(\frac{1}{5} , \frac{4}{5} , \frac{3}{5} ; 1-\eta )\, \Psi_{12,34} - F(\frac{1}{5} , - \frac{1}{5} , \frac{3}{5} ; 1-\eta )\, \Psi_{13,24}\,  ]\ ,\lb{Psi0-gen}\\
&&\T^{(1)}(\{w\}, \{ \uz \}) = C\, Q\,[\,(w_{12} w_{34})^{-3}\,\eta^3\, ] \left( w_{23} w_{14}\right)^{- \frac{4}{5}} \eta^{-\frac{7}{5}} (1-\eta)^{-\frac{3}{5}}\,\times \nn\\
&&\times \, [\, \eta\,F(\frac{6}{5} , \frac{3}{5} , \frac{7}{5} ; 1-\eta )\, \Psi_{12,34} + F(\frac{1}{5} , \frac{3}{5} , \frac{7}{5} ; 1-\eta )\, \Psi_{13,24}\,  ]\ .\lb{Psi1-gen}
\ea
Again, by (\ref{limPSI}) for $r=0\,$ (\ref{Psi0-gen}) and (\ref{Psi1-gen}) reproduce (\ref{Psi0-b2}) and (\ref{Psi1-b2}).
Rewriting the dual basis as
\ba
&&\T^{(0)}(\{w\}, \{ \uz \}) = Q\,[\,(w_{23} w_{14})^{-3}\,\eta^{-\frac{3}{2}}\,(1-\eta)^3\, ] \left( w_{23} w_{14}\right)^{- \frac{4}{5}} \eta^{\frac{1}{10}} (1-\eta)^{-1}\,\times \nn\\
&&\times \, [\, \eta\,F(\frac{1}{5} , \frac{4}{5} , \frac{3}{5} ; 1-\eta )\, \Psi_{12,34} - F(\frac{1}{5} , - \frac{1}{5} , \frac{3}{5} ; 1-\eta )\, \Psi_{13,24}\,  ]\ ,\lb{Psi0-gen1}\\
&&\T^{(1)}(\{w\}, \{ \uz \}) = C\, Q\,[\,(w_{23} w_{14})^{-3}\,\eta^{-\frac{3}{2}}\,(1-\eta)^3\, ] \left( w_{23} w_{14}\right)^{- \frac{4}{5}} \eta^{\frac{1}{10}} (1-\eta)^{-\frac{3}{5}}\,\times \nn\\
&&\times \, [\, \eta\,F(\frac{6}{5} , \frac{3}{5} , \frac{7}{5} ; 1-\eta )\, \Psi_{12,34} + F(\frac{1}{5} , \frac{3}{5} , \frac{7}{5} ; 1-\eta )\, \Psi_{13,24}\,  ]\lb{Psi1-gen1}
\ea
($(w_{23} w_{14})^{-3}\,\eta^{-\frac{3}{2}}\,(1-\eta)^3\,$ being invariant with respect to $B_2 :\, w_{23} \to e^{i\pi}\,w_{23}\,,$ cf. (\ref{B2eta})) and using
%again, as in the $r=0\,$ case,
(\ref{HG-ac11}) and (\ref{HGac5}) which imply
\ba
&&\xi^{(0)} := \eta\,F(\frac{1}{5} , \frac{4}{5} , \frac{3}{5} ; 1-\eta )\, \Psi_{12,34} - F(\frac{1}{5} , - \frac{1}{5} , \frac{3}{5} ; 1-\eta )\, \Psi_{13,24} \quad \to\quad -\, \eta^{-\frac{4}{5}}\,\xi^{(0)} \ ,\qquad\qquad\lb{B20}\\
&&\xi^{(1)} := \eta\,F(\frac{6}{5} , \frac{3}{5} , \frac{7}{5} ; 1-\eta )\, \Psi_{12,34} + F(\frac{1}{5} , \frac{3}{5} , \frac{7}{5} ; 1-\eta )\quad \to \quad
\eta^{-\frac{2}{5}}\, \xi^{(1)}\ ,\lb{B21}
\ea
we obtain that the counterparts of (\ref{Psi0-b}), (\ref{Psi1-b}) hold in the general case,
\be
b_2 \, \T(\{w\}, \{\uz\}) = R\,\, \T(\{w\}, \{\uz\})\,,\quad
\T(\{w\}, \{\uz\}) :=  \begin{pmatrix} \T^{(0)}(\{w\}, \{\uz\})\cr
\T^{(1)}( \{w\}, \{\uz\}) \end{pmatrix}
\lb{Psi-b-gen}
\ee
with the same matrix $R\,$ as in (\ref{BBRgen}). Hence, by (\ref{PhiFPsi-gen})
\be
b_2 \, \Phi(\{w\}, \{\uz\}) = B\, \Phi(\{w\}, \{\uz\})\ ,\quad  B = F\,R\,F \ ,\quad
F = \begin{pmatrix} \tau&\sqrt{\tau} \cr \sqrt{\tau}&- \tau\end{pmatrix} = F^{-1}\ .
\lb{b2Bgen}
\ee

The important conclusion that can be drawn from the computations in this section is that the presence of ($r\,$ triples of) electron fields doesn't change the braiding properties of the Fibonacci anyons, i.e. the latter are $r$-independent.

\bigskip

\noindent
{\bf Remark 3~} We recall that our primary object is the wave function (\ref{4eps}). The  proper braid matrices $B_i^{(4)}\,,\ i=1,2,3\,$ derived from it are obtained from (\ref{B3Phi}) by taking into account the additional Laughlin factors. 
%%$(w_i - w_{i+1})^{\frac{3}{5}}\,.$  
In effect, $B_i^{(4)} = q^3\, \pi^{(4)} (b_i)\,,$ or explicitly
\be
B_1^{(4)} = \begin{pmatrix} q^{-1} & 0 \cr 0 & - q \end{pmatrix}\ ,\quad
B_2^{(4)} = \begin{pmatrix} q^{-3} \tau &\sqrt{\tau} \cr \sqrt{\tau} & - q^3 \tau \end{pmatrix}\ ,\quad
B_3^{(4)} = B_1^{(4)}  \ .
\lb{Bq3}
\ee
The matrices (\ref{Bq3}) are the ones that are used in the paper \cite{GHM24}.

\section{Braiding a higher number of Fibonacci anyons}

\setcounter{equation}{0}
\renewcommand\theequation{\thesection.\arabic{equation}}

We will now propose a method which allows to formalize the procedure of finding the braidings of general $n$-point Fibonacci anyon conformal blocks. To this end, we first introduce, for $n\ge 4\,,$ the following notation for the vectors of the basis (corresponding to the admissible paths in the corresponding Bratteli diagram, see \cite{GHM24}):
\ba
&&\Phi^{0 \, 1\, \a_2 \, \a_3 \, \a_4 \, \dots\, \a_{n-2} \, 1 \, 0} = \Phi^{0 \, 1\, \a_2 \, \a_3 \, \a_4 \, \dots\, \a_{n-2} \, 1 \, 0} (\{w\}, \{ z\}) := \qquad\qquad\lb{1}\\
&&:= \langle 0 |\, \varepsilon (w_1) \Pi_1 \varepsilon (w_2) \Pi_{\a_2} \varepsilon (w_3) \Pi_{\a_3}\, \dots \, %\Pi_{\a_{n-3}}\, 
\varepsilon (w_{n-2}) \Pi_{\a_{n-2}} \varepsilon (w_{n-1}) \Pi_1\, \varepsilon (w_n)  \prod_{i=1}^{3r} \psi_1 (z_i) \, | 0 \rangle\ .\qquad
\nn
\ea
Here $\a_i\,,\ i = 2, \dots , \a_{n-2}\,$ take values $0\,$ or $1\,$
depending on whether the {\em orthogonal projector} $\Pi_{\a_i}\,$ projects on the vacuum or on the $\varepsilon^{( ' )} \,$ sector, respectively, cf. Remark 2 above; it is assumed that $\Pi_\a \Pi_\b = \d_{\a\b}\,\Pi_\a\,$ and $\Pi_0 + \Pi_1 = 1\,.$

Formula (\ref{1}) as it stays is relevant only for $n\,$ even, and when $n\,$ is odd, the fusion rules (\ref{eeOPE}), (\ref{Yee}) suggest that one (or, in general, an odd number) of the $\e\,$ fields should be replaced by $\e'\,.$ We will comment on this in more details in the discussion of the $n=3\,$ case below. Note that the $\psi_1\,$ fusion rules %, see e.g. \cite{AS07},
imply that $( \psi_1 )^3 \sim \id\,$ so that the product of $3\,r\,$ fields $\psi_1\,$ leaves the vacuum sector invariant; so does also the field $Y\,$ of (integer) dimension $3\,.$ A description of the sectors indexed by $0\,$ and $1\,$ (only a part of the full structure of $Z_3\,$ parafermion model) which is sufficient for our purposes is that $[ 0 ]\,$ contains vectors created from the vacuum by the Virasoro fields $\id\,$ and $Y\,,$ and the sector
$[ 1 ]\,,$ those created by $\e\,$ and $\e'\,.$ The fusion rules (\ref{Yee}) then imply
\be
[0] \times [0] = [0]\ ,\qquad [0] \times [1] \ (\, = [1] \times [0] \, )\ = [1]\ , \qquad [1] \times [1] = [0] \oplus [1 ]\ .
\lb{01}
\ee
So in the case of $n\,$ Fibonacci anyons the conformal blocks (\ref{1}) have $n+1\,$ indices altogether, the action of each Fibonacci field (from right to left) being specified by its initial and the target sector. It follows from (\ref{01}) that the (only) restriction of the ordered set $\a_2 \,, \a_3 \,, \a_4 \,, \dots\,, \a_{n-2}\,$ is that it should not contain two zero labels in a row.

The first and the last pair of indices ($01\,$ and $10\,,$ respectively) of $\Phi\,$ in (\ref{1}) are standard for the construction, and this fact does not leave room, when $n=2\,$ or $n=3\,$ for conformal blocks other than $\Phi^{010}\,$ and $\Phi^{0110}\,,$ respectively. (It complies, for $r = 0\,$ with the uniqueness, up to normalization,  of conformal invariant two- and three point functions.) In the first notrivial case $n=4\,$ the two possibilities $\Phi^{01010}\,$ and $\Phi^{01110}\,$ correspond to the conformal blocks of four Fibonacci anyons (\ref{4e-eta0}) and (\ref{4e-eta1}), respectively (or (\ref{4e-20}) and (\ref{4e-21}), in the $r=0\,$ case) and can be considered as a basis of the two dimensional representation of the braid group ${\cal B}_4\,$ generated by the matrices
%$\pi^{(4)} (b_1) = R = \pi^{(4)} (b_3)\ ,\ \pi^{(4)} (b_2) =
%B := F\,R\,F\ \ (F^2 = \id)\,,$
$R\,$ and $B := F R\,F\ \ (F^2 = \id)\,,$ see (\ref{BBRgen}) and (\ref{b2Bgen}).
As we are going to show below, this is sufficient to find the braid group ${\cal B}_n\,$ representation on the linear span $V_n\,$ of $n$-blocks of Fibonacci anyons (\ref{1}) for arbitrary $n\,.$

A simple observation suggests the following recursive construction.
Take $n\ge 4\,$ in (\ref{1}); now if the index $\a_{n-2}=0\,,$ then
$\Phi^{0 \, 1\, \a_2 \, \a_3 \, \a_4 \, \dots\, \a_{n-2}}\,$ could be any of the $(n-2)$-blocks, and if $\a_{n-2}=1\,,$  the vectors $\Phi^{0 \, 1\, \a_2 \, \a_3 \, \a_4 \, \dots\, \a_{n-2}\,0}\,$ span the space of $(n-1)$-blocks. The first two spaces $V_{d_2}\,$ and $V_{d_3}\,$ in this sequence are spanned by $\Phi^{010}\,$ and $\Phi^{0110}\,,$ respectively. Hence, for $n\ge 4\,$ any vector space $V_{d_n}\,$ is a direct sum and the dimensions $d_n = \dim V_{d_n}\,, \ n\ge 2\,$ form a Fibonacci sequence:
\be
V_{d_n} = V_{d_{n-2}} \oplus V_{d_{n-1}} \qquad \Rightarrow \qquad
d_n = d_{n-2} + d_{n-1}\qquad (d_n := \dim V_{d_n})\ ,\quad d_2 = d_3 = 1 \ .
\lb{Vn}
\ee
Accordingly, a basis in $V_{d_n}\,$ can be formed by taking first the vectors of the basis of $V_{d_{n-2}}\,$ (just replacing their last indices $1 0\,$ by $ 1 0 1 0$) and then those of $V_{d_{n-1}}\,,$ replacing this time $1 0\,$ by  $ 1 1 0\,;$ we will also assume that the internal ordering of the bases of the subspaces is inherited.
The braiding for $n=2\,$ (and $r=0$) follows from the two point function
\be
\langle\, \e (w_1) \, \e (w_2) \rangle = w_{12}^{-\frac{4}{5}}\ ,\quad
w_{12} \to e^{i\pi} w_{12} \quad\Rightarrow\quad b_1 \,\Phi^{010} = q^{-4}\, \Phi^{010}\qquad (q= e^{i \frac{\pi}{5}})\ .
\lb{n2}
\ee
The next value $n=3\,$ being odd, we must replace one of the $\e\,$ fields by $\e'\,.$ Choosing this to be the last one in the three point function, we obtain
\be
\langle\, \e (w_1) \, \e (w_2) \, \e' (w_3) \rangle = C_{\e' \e \e}\, w_{12}^{\frac{3}{5}}\, (w_{13}\, w_{23})^{-\frac{7}{5}} \quad\Rightarrow\quad b_1 \Phi^{0110} = q^3\, \Phi^{0110}\ .
\lb{n3}
\ee
To compute $b_2\,,$ we compare (\ref{n3}) with
\be
\langle\, \e (w_1) \, \e' (w_3) \, \e (w_2) \rangle = C_{\e' \e \e}\, w_{12}^{\frac{3}{5}}\, (w_{13}\, w_{32})^{-\frac{7}{5}}\ ,\quad w_{32} = e^{i\pi} w_{23} \quad\Rightarrow\quad b_2\, \Phi^{0110} = q^3\, \Phi^{0110}
\lb{n32}
\ee
(see Remark 1 above), using also that $q^{10} = 1\,$ and hence, $q^{-7} = q^3\,;$
note that we exchange the (posititions of the) fields, not just their arguments. To summarize,
\be
\pi^{(2)} (b_1) = q^{-4}\ ,\qquad \pi^{(3)} (b_1) = q^3 = \pi^{(3)} (b_2)\ .
\lb{n23}
\ee
As the representation of the braid group ${\cal B}_3\,$ is one dimensional, the Artin relation $\pi^{(3)} (b_1\, b_2\, b_1) = \pi^{(3)} (b_2\, b_1\, b_2 )\,$ is trivially satisfied. It is a simple exercise to show that the same results are obtained starting with any other position of $\e'\,$ in the three point function or even with the correlator of three $\e'\,$ fields,
\ba
\langle\, \e' (w_1) \, \e (w_2) \, \e (w_3) \rangle &=& C_{\e' \e \e}\,w_{23}^{\frac{3}{5}}\, (w_{12}\, w_{13})^{-\frac{7}{5}} \ ,\nn\\
\langle\, \e' (w_1) \, \e' (w_2) \, \e' (w_3) \rangle
&=& C_{\e' \e' \e'}\,(w_{12}\,w_{13}\,w_{23})^{-\frac{7}{5}}
\lb{n3var}
\ea
(or, in the $n=2\,$ case, from $\langle\, \e' (w_1) \, \e' (w_2) \rangle = z_{12}^{-\frac{14}{5}}$).

In general, the braid group generator $\pi^{(n)} (b_i)\,$ acting on the {\em $i$-th triple of consecutive indices} ${\a_{i-1}}\, {\a_i}\,{\a_{i+1}}\,$ of the vector $\Phi^{\,\dots\, {\a_{i-1}}\, {\a_i}\,{\a_{i+1}} \dots}\,$ (\ref{1}) corresponds to the exchange of $\varepsilon (w_i)\,$ and $\varepsilon (w_{i+1})\,$ along certain classes of paths not enclosing any of the other points. The rules (\ref{01}) suggest that these vectors form singlets when either $\a_{i-1}\,$ or $\a_{i+1}\,,$ or both, are zero (then $\a_i\,$ can only be equal to $1$), and from (\ref{n2}), (\ref{n3}) and (\ref{n32}) one would expect that
\ba
&&\pi^{(n)} (b_i) \, \Phi^{\dots 010 \dots} = q^{-4} \,\Phi^{\dots 010 \dots}\ ,\nn\\
&&\pi^{(n)} (b_i) \, \Phi^{\dots 011 \dots} = q^3 \,\Phi^{\dots 011 \dots}\ ,\quad
\pi^{(n)} (b_i) \, \Phi^{\dots 110 \dots} = q^3 \,\Phi^{\dots 110 \dots}\ .
\lb{2-3}
\ea
This is confirmed by the results for the four-point blocks written as $\Phi^{01010}\,$ and $\Phi^{01110}\,,$ respectively where we have (see (\ref{B3Phi}) for the $r=0\,$ case)
\be
b_1 \begin{pmatrix} \Phi^{01010} \cr \Phi^{01110} \end{pmatrix} = R\, \begin{pmatrix} \Phi^{01010} \cr \Phi^{01110} \end{pmatrix}\ ,\quad
R = \pi^{(4)} (b_1) = \begin{pmatrix} q^{-4}&0\cr 0& q^3\end{pmatrix} = \begin{pmatrix} \pi^{(2)} (b_1) &0\cr 0& \pi^{(3)} (b_1)\end{pmatrix}
\lb{n4-12}
\ee
and similarly, $\pi^{(4)} (b_3) = R\,.$ (In both cases the action of the braidings on the corresponding singlets is combined in the diagonal matrix $R\,.$) 
On the other hand, the $b_2\,$ action (\ref{B3Phi}) can be written as
\be
b_2 \begin{pmatrix} \Phi^{01010} \cr \Phi^{01110} \end{pmatrix} = B\, \begin{pmatrix} \Phi^{01010} \cr \Phi^{01110} \end{pmatrix}\ ,\quad
B = \pi^{(4)} (b_2) = \begin{pmatrix} B^0_{~0}&B^0_{~1} \cr B^1_{~0}& B^1_{~1} \end{pmatrix} = \begin{pmatrix} q^4 \tau & q^{-3} \sqrt{\tau} \cr q^{-3} \sqrt{\tau} & - \tau\end{pmatrix}\ ,
\lb{n4-23}
\ee
suggesting that for $\a_{i-1} = 1 = \a_{i+1}\,$ the braiding 
$\pi^{(n)} (b_i)\,$ acts on doublets (since in this case $\a_i\,$ can be $0\,$ or $1$), and
\be
\pi^{(n)} (b_i) \begin{pmatrix} \Phi^{\dots 101 \dots} \cr \Phi^{\dots 111 \dots} \end{pmatrix} = B\, \begin{pmatrix} \Phi^{\dots 101 \dots} \cr \Phi^{\dots 111 \dots} \end{pmatrix}\ .
\lb{nn-23}
\ee
(In (\ref{nn-23}) we assume that all other indices of $\Phi^{\dots 101 \dots}\,$  coincide with those of $\Phi^{\dots 111 \dots}\,.$)

In the next section we will provide a general argument why the same (diagonal elements of) $R\,$ and the $2 \times 2\,$ matrix $B\,$ derived from the two- and three-point anyon functions and the four-point conformal blocks should appear as matrix blocks in the higher $n\,$ braiding matrices. We will then use the algorithm described above to obtain an explicit recursive construction of the braid group ${\cal B}_n\,$ action on $n\,$ Fibonacci anyons for any $n\,.$

\medskip

\noindent
{\bf Remark 4~} Note that the linear algebra prescription for assigning a matrix to an operator in a given basis would require to take actually the {\em transposed} of the (non-diagonal) matrices. The (wrong) traditional definition finds a partial excuse in the fact that all Artin relations (like (\ref{ArtinB4})) are invariant with respect to matrix transposition.

\medskip

\section{Explicit recursive construction of the ${\cal B}_n$ action for $n \ge 5$}  %small 

\setcounter{equation}{0}
\renewcommand\theequation{\thesection.\arabic{equation}}

We will begin this section by recalling that the abstract Artin braid group ${\cal B}_n\,$ generated by $n-1\,$ generators $b_i\,$ satisfying
\be
b_i \, b_{i+1}\, b_i = b_{i+1}\, b_i \, b_{i+1}\ ,\quad i=1, \dots, n-2\ ,\qquad b_i \,b_j = b_j\, b_i\ ,\quad |i-j|\ge 2\ 
\lb{BArt}
\ee
is the proper algebraic structure to handle generalized statistics 
in low dimensional physics (\cite{S23}; for a concise introduction to the subject, see e.g. \cite{TH01}). As the configuration space of   $n\,$ points on a two dimensional surface is not simply connected, the corresponding wave function depending on $n\,$ complex variables may be multivalued. The latter is a characteristic property of the quasiparticles called "anyons" by F. Wilczek in 1982. 

It is obvious from (\ref{BArt}) that a natural sequence of braid group inclusions
\be
{\cal B}_2 \subset {\cal B}_3 \subset \dots \subset {\cal B}_{n-1} \subset {\cal B}_n
\lb{Bn-incl}
\ee
exists, each of the subgroups ${\cal B}_i\,$ of ${\cal B}_n\,,\ 2\le i \le n\,$ in (\ref{Bn-incl}) being generated by the first $i-1\,$ generators $b_j\,,\ j=1, \dots , i-1\,.$ The latter assumption (concerning the identification of the subgroups) is actually conventional: one can start the sequence (\ref{Bn-incl}) e.g. with $b_{n-1}\,$ (generating a ${\cal B}_2\,$ subgroup) and proceed by including consecutive generators with smaller indices, $b_j\,,\ j=n-1, n-2, \dots ,1\,.$ 

In our case the generators $b_i\,$ of the braid group 
%% ${\cal B}_n\,$ 
correspond to the exchange of neighboring Fibonacci fields along certain (homotopy classes of) paths so that e.g.
$b_i \equiv b_{i\,i+1}: \ w_{i\,i+1} \stackrel{\curvearrowleft}{\longrightarrow} w_{i+1\, i} := e^{i\pi} w_{i\,i+1}\,.$ This induces, equivalently, a linear transformation on the $n$-point (in $\{ w \})\,$ conformal blocks defining the "monodromy representation" of ${\cal B}_n\,.$ We denote by $\pi^{(n)} (b_i)\,$ the corresponding matrix in the basis (\ref{1}). The calculation of the braid matrices has been carried out in detail in the previous sections in the cases $n \le 4\,$ (being only non-trivial for $n=4\,,$ of course) using the explicit form of the relevant correlators.

At first sight, extending the braiding action to cases with higher number of Fibonacci anyons $n>4\,$ could be difficult as the  corresponding anyon correlators are not known explicitly. To this end, however, one can use the "locality" of the action of Artin braid group generators in the sense that they only affect the positions of two neighboring points (anyon coordinates) while the rest of the anyons play the role of spectators. This observation allows us to use the (short distance) OPE (\ref{Yee}) to reduce the number of Fibonacci fields in the $n$-point conformal blocks. We will sketch in what follows the main steps of this procedure for $r=0\,$ (which is not a restriction as we know that the result, what concerns braiding, doesn't depend on $r\,$), starting for concreteness with $n\,$ even. Applying (\ref{Yee}) to the last two Fibonacci fields in (\ref{1}) in this case, we obtain that, for $w_{n-1} \sim w_n\,$
\ba
&&\varepsilon (w_{n-2})\,
\Pi_{\a_{n-2}} \,\varepsilon (w_{n-1}) \,\Pi_1\, \varepsilon (w_n)  \, 
| 0 \rangle = \varepsilon (w_{n-2})\,
\Pi_{\a_{n-2}} \,\varepsilon (w_{n-1}) \, \varepsilon (w_n)  \, 
| 0 \rangle \ \sim\nn\\
&&\sim\, \varepsilon (w_{n-2})\,
\Pi_{\a_{n-2}} \left( (w_{n-1}-w_n)^{-\frac{4}{5}} \,\id + 
\sqrt{\frac{12}{7}\, C}\, (w_{n-1}-w_n)^{\frac{3}{5}} \,\e'(w_n) \right) | 0 \rangle =
\nn\\
&& = \left\{
\begin{array}{ll}
\, (w_{n-1}-w_n)^{-\frac{4}{5}} \, \varepsilon (w_{n-2})\,| 0 \rangle &, \quad \a_{n-2} = 0\\
\, \sqrt{\frac{12}{7}\, C}\,(w_{n-1}-w_n)^{\frac{3}{5}} \, \varepsilon (w_{n-2})\, \Pi_1 \, \e' (w_{n})\, | 0 \rangle &, \quad \a_{n-2} = 1
\end{array}
\right.\quad \ ,
\lb{arg1}
\ea
see (\ref{eeOPE}) and (\ref{Ceee}).
As we know, for $n\,$ odd we should have an odd number of $\e'\,$ fields in the correlator. To sketch the needed modification of (\ref{arg1}) we will write, for example, schematically
\be 
\e \, \e \, \e' \, | 0 \rangle \ \sim \
\e \, (Y + \e )\, | 0 \rangle 
\ \sim \ \e' \, | 0 \rangle +  \e \, \e \, | 0 \rangle \ 
\lb{arg1odd}
\ee
see (\ref{Yee}) in Remark 1.

Inserting (\ref{arg1}) (resp. (\ref{arg1odd}), for $n\,$ odd) into (\ref{1}) for $r=0\,$ we express, for $w_{n-1} \sim w_n\,,$ $n$-point conformal blocks as sums of $(n-2)$-point and $(n-1)$-point ones. Accordingly, we can compute the first $n-3\,$ braiding matrices $\pi^{(n)} (b_i)\,,\ i = 1, 2,\dots, n-3\,$ (generating the subgroup   ${\cal B}_{n-2}\,$) from those for $\pi^{(n-2)} (b_i)\,$ and $\pi^{(n-1)} (b_i)\,. $ Moreover, the description given in the paragraph after (\ref{Vn}) of the construction of the $V_n\,$ basis i.e., writing first the vector components inherited from $V_{n-2}\,$ (those with last three indices equal to $0 1 0\,$) followed by those from $V_{n-1}\,$ (with last three indices $1 1 0\,$) anticipates as well the block diagonal form of the matrices of these braid generators encountered in the explicit calculations, the results of which are displayed below.

The above reduction procedure becomes effective for $n \ge 5\,.$ Obviously, it cannot be used directly for the derivation of the last two braidings, $\pi^{(n)} (b_{n-2})\,$ and  $\pi^{(n)} (b_{n-1})\,$ 
(in accord with the fact that one, or two of the terms in the recursion simply do not exist in these cases). Then it can be replaced, however, by a similar procedure involving the left vacuum and the first three Fibonacci fields in place of (\ref{arg1}) or (\ref{arg1odd}). Thus, one can recover this time {\it the last} $n-3\,$ generators of ${\cal B}_n\,$ in terms of those of ${\cal B}_{n-2}\,$ and ${\cal B}_{n-1}\,.$ 

In particular, the matrix $\pi^{(n)} (b_{n-2})\,$ is expressible through $\pi^{(n-2)} (b_{n-4})\,$ and $\pi^{(n-1)} (b_{n-3})\,.$ One can anticipate that it wouldn't have block diagonal structure in the basis (\ref{1}) in which the branches of the "recursion tree" of $V_n\,$ subspaces are determined, at every step, by the last triple of indices (corresponding to the two possible fusion channels of the last two Fibonacci anyons). In contrast, the OPE of the first two Fibonacci anyons would correspond to a different subspace decomposition depending on the first triple of indices (either $0 1 0 \,$ or $0 1 1\,$). 

We will illustrate this in the simplest, $n=5\,$ case. To obtain a basis in $V_{d_5}\,$ we present, according to our convention, the three possible conformal blocks as a vector with a single "upper" and two "lower" components
\ba
&&\begin{pmatrix} \Phi^{011010}\cr \Phi^{010110}\cr \Phi^{011110}\end{pmatrix} = \begin{pmatrix} \Phi^{011010}\cr 0 \cr 0\end{pmatrix} \ \oplus\ 
\begin{pmatrix} 0 \cr \Phi^{010110} \cr \Phi^{011110}\end{pmatrix} \ ,\nn\\
&&V_{d_5} \simeq V_{d_3} \oplus V_{d_4}\ , \ d_5 = d_3 + d_4 = 1 + 2 = 3 
\lb{5pt'}
\ea
(see (\ref{Vn})), the decomposition matching the fusion of the fourth and fifth Fibonacci anyons in the $5$-point conformal blocks (\ref{1}). For comparison, the fusion of the first two anyons corresponds to a subspace decomposition of the type
\be
\begin{pmatrix} \Phi^{011010}\cr \Phi^{010110}\cr \Phi^{011110}\end{pmatrix} = 
\begin{pmatrix} 0 \cr \Phi^{010110}\cr 0\end{pmatrix} \ \oplus\ 
\begin{pmatrix} \Phi^{011010}\cr 0 \cr \Phi^{011110}\end{pmatrix} \ ,
\lb{leftOPE}
\ee
so that $\pi^{(5)} (b_3)\,$ decomposes into $\pi^{(3)} (b_1)\ (= q^3\,,$ cf. (\ref{n23})) acting on the singlet and $\pi^{(4)} (b_2)\ (= B\,,$ see (\ref{n4-23})) acting on the doublet in (\ref{leftOPE}), accordingly. In general, $\pi^{(n)} (b_{n-2})\,$ has again a direct sum structure (but is block diagonal in a basis different from the "canonical" one).

It is easy to realize that the matrices $\pi^{(n)} (b_{n-1})\,$ and  $\pi^{(n)} (b_1)\,$ should be diagonal for all $n\,;$ in particular, $\pi^{(5)} (b_4)\,$ acts again by multiplication by $q^3 = \pi^{(3)} (b_2)\,$ on the singlet, and by the diagonal matrix $R = \pi^{(4)}(b_3)\,$ on the doublet in (\ref{leftOPE}). 

An interesting phenomenon (providing a self-consistency check of the above OPE reduction procedure) appears for the first time in the next example $n=6\,,$ where the braiding matrix $\pi^{(6)} (b_3)\,$
can be computed in two different ways. The first one feflects its "canonical" block diagonal decomposition into $\pi^{(4)} (b_3)\,$ and $\pi^{(5)} (b_3)\,$ and the other (the reduction implied by fusing the first two Fibonacci anyons instead of the last two ones), from   $\pi^{(4)} (b_1)\ (= R\,,$ see (\ref{n4-12}), acting on the first and the fourth component subspace of $V_6\,$) and $\pi^{(5)} (b_2)\,,$ cf. (\ref{B5}) and (\ref{B6}) below.

The most important conclusion from the above consideration is that, going backwards (in $n\,$), it is possible indeed to recover the braiding action for any $n\,$ from the $2$-, $3$- and $4$-point braiding data. 

To verify that the OPE reduction procedure reproduces the results following from the prescriptions described in the previous section, we will use the latter to compute the braiding matrices for $n=5 , 6 , 7\,$ and $8\,.$ As explained above, the generator $b_i\,$ of the Artin braid group ${\cal B}_n\,$ (for any $i = 1, \dots , n-1$) is represented by the matrix $\pi^{(n)}(b_i)\,$ obtained by applying specific rules to the $i$-th consecutive triple of indices of the vectors in (\ref{1}). There are $n+1\,$ indices altogether in $n$-blocks so that the number of consecutive triples is  $n-1\,,$ matching the number of generators of ${\cal B}_n\,.$ The corresponding action on a specific triple is given by (\ref{2-3}) (multiplication by a phase, $q^{-4}\,$ or $q^3$) or, for doublets of the form (\ref{n4-23}), by the $2\times 2\,$ matrix
\begin{equation*}
B = \begin{pmatrix} B^0_{~0}&B^0_{~1} \cr B^1_{~0}& B^1_{~1} \end{pmatrix} = \begin{pmatrix} q^4 \tau & q^{-3} \sqrt{\tau} \cr q^{-3} \sqrt{\tau} & - \tau\end{pmatrix}\ .
\end{equation*}

Following the rules and conventions spelled out above, we can compute the matrices of the four generators of ${\cal B}_5\,$ in the basis (\ref{5pt'}). They are given by
\ba
&&\pi^{(5)} (b_1) =
\begin{pmatrix}
q^3 & 0 & 0 \cr
0 & q^{-4} & 0 \cr
0 & 0 & q^3
\end{pmatrix}\ ,\qquad\ \,
\pi^{(5)} (b_2) =
\begin{pmatrix}
q^3 & 0 & 0 \cr
0 & B^0_{~0} & B^0_{~1} \cr
0 & B^1_{~0} & B^1_{~1}
\end{pmatrix}\ ,\nn\\
&&\pi^{(5)} (b_3) =
\begin{pmatrix}
B^0_{~0} & 0 & B^0_{~1} \cr
0 & q^3 & 0 \cr
B^1_{~0} & 0 & B^1_{~1}
\end{pmatrix}\ ,\qquad
\pi^{(5)} (b_4) =
\begin{pmatrix}
q^{-4} & 0 & 0 \cr
0 & q^3 & 0 \cr
0 & 0 & q^3
\end{pmatrix}\ .
\lb{B5}
\ea
Almost all of the Artin relations (\ref{BArt}) are easy to verify in the $n=5\,$ case (\ref{B5}). First of all, the commutators $[\pi^{(5)}(b_1) , \pi^{(5)}(b_3)]\,$ and $[\pi^{(5)}(b_2) , \pi^{(5)}(b_4)]\,$ vanish because of the matching block structure in which the counterparts of the $2\times 2\,$ non-diagonal submatrices are proportional to the unit $2\times 2\,$ matrix, and $[\pi^{(5)}(b_1) , \pi^{(5)}(b_4) ] = 0\,$ since both matrices are diagonal. One observes further that the triple product Artin relations $\pi^{(5)}(b_1\, b_2\, b_1) = \pi^{(5)}(b_2\, b_1\, b_2)\,$ and $\pi^{(5)}(b_3\, b_4\, b_3) = \pi^{(5)}(b_4\, b_3\, b_4)\,$ follow, essentially, from the basic $2 \times 2\,$ relation $R\, B\, R = B\, R\, B\,$ (\ref{Artin}).

The only non-trivial relation that remains to be verified is therefore the equality of $\pi^{(5)}(b_2\, b_3\, b_2) = \pi^{(5)}(b_3\, b_2\, b_3)\,.$ From (\ref{B5}) we obtain
\ba
&&\pi^{(5)} (b_2\, b_3\, b_2) =
\begin{pmatrix}
q^6 B^0_{~0} & q^3 B^0_{~1} B^1_{~0} & q^3 B^0_{~1} B^1_{~1} \cr
q^3 B^0_{~1} B^1_{~0}\ &\ q^3 (B^0_{~0})^2 + B^0_{~1} B^1_{~0} B^1_{~1}\ &\ B^0_{~1} (q^3 B^0_{~0} + (B^1_{~1})^2)\ \cr
q^3 B^1_{~1} B^1_{~0} & B^1_{~0} (q^3 B^0_{~0} + (B^1_{~1})^2) & q^3 B^0_{~1} B^1_{~0} + (B^1_{~1})^3
\end{pmatrix}\ ,\nn\\
&&\pi^{(5)} (b_3\, b_2\, b_3) =
\begin{pmatrix}
q^3 (B^0_{~0})^2 + B^0_{~1} B^1_{~0} B^1_{~1}\ &\ q^3 B^0_{~1} B^1_{~0}\ &\ B^0_{~1} (q^3 B^0_{~0} + (B^1_{~1})^2) \cr
q^3 B^0_{~1} B^1_{~0}  & q^6 B^0_{~0} & q^3 B^0_{~1} B^1_{~1} \cr
B^1_{~0} (q^3 B^0_{~0} + (B^1_{~1})^2) & q^3 B^1_{~0} B^1_{~1} & q^3 B^0_{~1} B^1_{~0} + (B^1_{~1})^3
\end{pmatrix}\ .\qquad
\lb{checkb2b3}
\ea
Inserting the actual values of the entries of the $B\,$ matrix, % from (\ref{B_2}),
we get
\ba
&&\pi^{(5)} (b_2\, b_3\, b_2) =
\begin{pmatrix}
\tau & q^{-3}\tau & - \tau \sqrt{\tau}\ \cr
q^{-3}\tau & \tau & (q^4 + q^{-3} \tau ) \tau \sqrt{\tau}\cr
- \tau \sqrt{\tau}\ & \ (q^4 + q^{-3} \tau ) \tau \sqrt{\tau} & (q^{-3} - \tau^2)\tau
\end{pmatrix}\ ,\nn\\
&&\pi^{(5)} (b_3\, b_2\, b_3) =
\begin{pmatrix}
(q+q^{-1}) \tau^2  &\  q^{-3}\tau \ &\  (q^4 + q^{-3} \tau ) \tau \sqrt{\tau} \cr
q^{-3}\tau  & \tau & - \tau \sqrt{\tau} \cr
(q^4 + q^{-3} \tau ) \tau \sqrt{\tau}\ &\ - \tau \sqrt{\tau}  & (q^{-3} - \tau^2)\tau
\end{pmatrix}\ .\qquad
\lb{checkb2b3-1}
\ea
It remains to use $q^4 + q^{-3} \tau = q^4 + q^{-1} + q^{-5} = - 1\,$ for $\tau = \frac{1}{q+q^{-1}} =          q^2 + q^{-2}\,,$ cf. (\ref{tau}).

\vspace{5mm}

Following the prescriptions, we arrange the five basis vectors for the $n=6\,$ Fibonacci conformal blocks in the following order:
\be
\begin{pmatrix} \Phi^{0101010}\cr \Phi^{0111010}\cr \Phi^{0110110}\cr \Phi^{0101110} \cr \Phi^{0111110}\end{pmatrix}\qquad ( V_{d_6} = V_{d_4} \oplus V_{d_5} \ ,\ d_6 = d_4 + d_5 = 2 + 3 = 5)\ .
\lb{6pt1}
\ee
The matrices of the braid group ${\cal B}_6\,$ generators are then given by
\ba
&&\pi^{(6)} (b_1) = \begin{pmatrix}
q^{-4} & 0 & 0 & 0 & 0 \cr
0 & q^3 & 0 & 0 & 0 \cr
0 & 0 & q^3 & 0 & 0 \cr
0 & 0 & 0 & q^{-4} & 0 \cr
0 & 0 & 0 & 0 & q^3
\end{pmatrix}\ , \quad\quad
\pi^{(6)} (b_2) = \begin{pmatrix}
B^0_{~0} & B^0_{~1} & 0 & 0 & 0 \cr
B^1_{~0} & B^1_{~1} & 0 & 0 & 0 \cr
0 & 0 & q^3 & 0 & 0 \cr
0 & 0 & 0 & B^0_{~0} & B^0_{~1}\cr
0 & 0 & 0 & B^1_{~0} & B^1_{~1}
\end{pmatrix}\ , \nn\\
&&\pi^{(6)} (b_3) = \begin{pmatrix}
q^{-4} & 0 & 0 & 0 & 0 \cr
0 & q^3 & 0 & 0 & 0 \cr
0 & 0 & B^0_{~0} & 0 & B^0_{~1} \cr
0 & 0 & 0 & q^3 & 0 \cr
0 & 0 & B^1_{~0} & 0 & B^1_{~1}
\end{pmatrix}\ , \quad
\pi^{(6)} (b_4) = \begin{pmatrix}
B^0_{~0} & 0 & 0 & B^0_{~1} & 0 \cr
0& B^0_{~0} & 0 & 0 &  B^0_{~1}  \cr
0 & 0 & q^3 & 0 & 0 \cr
B^1_{~0}& 0 & 0 & B^1_{~1} & 0\cr
0 & B^1_{~0} & 0 & 0 & B^1_{~1}
\end{pmatrix}\ , \nn\\
&&\pi^{(6)} (b_5) = \begin{pmatrix}
q^{-4} & 0 & 0 & 0 & 0 \cr
0 & q^{-4} & 0 & 0 & 0 \cr
0 & 0 & q^3 & 0 & 0 \cr
0 & 0 & 0 & q^3 & 0 \cr
0 & 0 & 0 & 0 & q^3
\end{pmatrix}\ .\qquad
\lb{B6}
\ea
Verifying the Artin relations by hand as in the previous, $n=5\,$ case is still feasible (note again the block submatrix structure which could be helpful in some calculations) but tedious so we will skip it here. The triple Artin relations in (\ref{Artin}) require the determinants of all $\pi^{(n)}(b_i)\,$ in a given representation to be equal. Denoting their common value by $D_n\,,$ we see that $D_4 = q^{-1}\,$ (cf. (\ref{dets})) while $D_5 = q^2\,$ and $D_6 = q\,.$

\smallskip

We will also display the results for the next two representations, of dimensions $d_7 = 3 + 5 = 8\,$ and $d_8 = 5 + 8 = 13\,$ which are not too hard to obtain even without any computer help.

\newpage

\noindent
{\bf Basis vectors and braid matrices for $n=7$}

\medskip

\noindent
{\bf Basis of} $V_{d_7}$:

\be
\begin{pmatrix}
\Phi^{01101010}\cr \Phi^{01011010}\cr \Phi^{01111010}\cr
\Phi^{01010110}\cr \Phi^{01110110}\cr \Phi^{01101110}\cr \Phi^{01011110} \cr \Phi^{01111110}\end{pmatrix}
\qquad ( V_{d_7} = V_{d_5} \oplus V_{d_6} \ ,\ d_7 = d_5 + d_6 = 3 + 5 = 8)
\lb{7pt}
\ee

\medskip

\noindent
{\bf ${\cal B}_7\,$ generators} $\pi^{(7)}(b_i) \,,\ i = 1,\dots, 6\,,\quad D_7 = q^3$:

\ba
&&\pi^{(7)}(b_1) = \begin{pmatrix}
q^3 & 0 & 0 & 0 & 0 & 0 & 0 & 0 \cr
0 & q^{-4} & 0 & 0 & 0 & 0 & 0 & 0 \cr
0 & 0 & q^3 & 0 & 0 & 0 & 0 & 0 \cr
0 & 0 & 0 & q^{-4} & 0 & 0 & 0 & 0 \cr
0 & 0 & 0 & 0 & q^3 & 0 & 0 & 0 \cr
0 & 0 & 0 & 0 & 0 & q^3 & 0 & 0 \cr
0 & 0 & 0 & 0 & 0 & 0 & q^{-4} & 0 \cr
0 & 0 & 0 & 0 & 0 & 0 & 0 & q^3
\end{pmatrix}\ ,\nn
\ea

\ba
&&\pi^{(7)}(b_2) = \begin{pmatrix}
q^3 & 0 & 0 & 0 & 0 & 0 & 0 & 0 \cr
0 & B^0_{~0} & B^0_{~1} & 0 & 0 & 0 & 0 & 0 \cr
0 & B^1_{~0} & B^1_{~1} & 0 & 0 & 0 & 0 & 0 \cr
0 & 0 & 0 & B^0_{~0} & B^0_{~1} & 0 & 0 & 0 \cr
0 & 0 & 0 & B^1_{~0} & B^1_{~1} & 0 & 0 & 0 \cr
0 & 0 & 0 & 0 & 0 & q^3 & 0 & 0 \cr
0 & 0 & 0 & 0 & 0 & 0 & B^0_{~0} & B^0_{~1} \cr
0 & 0 & 0 & 0 & 0 & 0 & B^1_{~0} & B^1_{~1}
\end{pmatrix}\ ,\nn
\ea

\ba
&&\pi^{(7)}(b_3) = \begin{pmatrix}
B^0_{~0} & 0 & B^0_{~1} & 0 & 0 & 0 & 0 & 0 \cr
0 & q^3 & 0 & 0 & 0 & 0 & 0 & 0 \cr
B^1_{~0} & 0 & B^1_{~1} & 0 & 0 & 0 & 0 & 0 \cr
0 & 0 & 0 & q^{-4} & 0 & 0 & 0 & 0 \cr
0 & 0 & 0 & 0 & q^3 & 0 & 0 & 0 \cr
0 & 0 & 0 & 0 & 0 & B^0_{~0} & 0 & B^0_{~1} \cr
0 & 0 & 0 & 0 & 0 & 0 & q^3 & 0 \cr
0 & 0 & 0 & 0 & 0 & B^1_{~0} & 0 & B^1_{~1}
\end{pmatrix} \ ,\nn
\ea

\ba
&&\pi^{(7)}(b_4) = \begin{pmatrix}
q^{-4} & 0 & 0 & 0 & 0 & 0 & 0 & 0 \cr
0 & q^3 & 0 & 0 & 0 & 0 & 0 & 0 \cr
0 & 0 & q^3 & 0 & 0 & 0 & 0 & 0 \cr
0 & 0 & 0 & B^0_{~0} & 0 & 0 & B^0_{~1} & 0 \cr
0 & 0 & 0 & 0 & B^0_{~0} & 0 & 0 & B^0_{~1} \cr
0 & 0 & 0 & 0 & 0 & q^3 & 0 & 0 \cr
0 & 0 & 0 & B^1_{~0} & 0 & 0 & B^1_{~1} & 0 \cr
0 & 0 & 0 & 0 & B^1_{~0} & 0 & 0 & B^1_{~1}
\end{pmatrix}\ ,\nn
\ea

\ba
&&\pi^{(7)}(b_5) = \begin{pmatrix}
B^0_{~0} & 0 & 0 & 0 & 0 & B^0_{~1} & 0 & 0 \cr
0 & B^0_{~0} & 0 & 0 & 0 & 0 & B^0_{~1} & 0 \cr
0 & 0 & B^0_{~0} & 0 & 0 & 0 & 0 & B^0_{~1} \cr
0 & 0 & 0 & q^3 & 0 & 0 & 0 & 0 \cr
0 & 0 & 0 & 0 & q^3 & 0 & 0 & 0 \cr
B^1_{~0} & 0 & 0 & 0 & 0 & B^1_{~1} & 0 & 0 \cr
0 & B^1_{~0} & 0 & 0 & 0 & 0 & B^1_{~1} & 0 \cr
0 & 0 & B^1_{~0} & 0 & 0 & 0 & 0 & B^1_{~1}
\end{pmatrix} \ ,\nn
\ea

\ba
&&\pi^{(7)}(b_6) = \begin{pmatrix}
q^{-4} & 0 & 0 & 0 & 0 & 0 & 0 & 0 \cr
0 & q^{-4} & 0 & 0 & 0 & 0 & 0 & 0 \cr
0 & 0 & q^{-4} & 0 & 0 & 0 & 0 & 0 \cr
0 & 0 & 0 & q^3 & 0 & 0 & 0 & 0 \cr
0 & 0 & 0 & 0 & q^3 & 0 & 0 & 0 \cr
0 & 0 & 0 & 0 & 0 & q^3 & 0 & 0 \cr
0 & 0 & 0 & 0 & 0 & 0 & q^3 & 0 \cr
0 & 0 & 0 & 0 & 0 & 0 & 0 & q^3
\end{pmatrix}\ , \lb{B7}
\ea

\newpage

\noindent
{\bf Basis vectors and braid matrices for $n=8$}

\medskip

\noindent
{\bf Basis of} $V_{d_8}\,$:

\be
\begin{pmatrix} \Phi^{010101010}\cr \Phi^{011101010}\cr \Phi^{011011010}\cr \Phi^{010111010} \cr \Phi^{011111010}\cr
\Phi^{011010110}\cr \Phi^{010110110}\cr \Phi^{011110110}\cr
\Phi^{010101110}\cr \Phi^{011101110}\cr \Phi^{011011110}\cr \Phi^{010111110} \cr \Phi^{011111110}\end{pmatrix}
\qquad ( V_{d_8} = V_{d_6} \oplus V_{d_7} \ ,\ d_8 = d_6 + d_7 = 5 + 8 = 13)
\lb{8pt}
\ee

\medskip

\noindent
{\bf ${\cal B}_8\,$ generators} $\pi^{(8)}(b_i) \,,\ i = 1,\dots, 7\,,\quad D_8 = - q^{-1}$:

\setcounter{MaxMatrixCols}{20}

\ba
\pi^{(8)}(b_1) = \begin{pmatrix}
q^{-4} & 0 & 0 & 0 & 0 & 0 & 0 & 0 & 0 & 0 & 0 & 0 & 0 \cr
0 & q^3 & 0 & 0 & 0 & 0 & 0 & 0 & 0 & 0 & 0 & 0 & 0 \cr
0 & 0 & q^3 & 0 & 0 & 0 & 0 & 0 & 0 & 0 & 0 & 0 & 0 \cr
0 & 0 & 0 & q^{-4} & 0 & 0 & 0 & 0 & 0 & 0 & 0 & 0 & 0 \cr
0 & 0 & 0 & 0 & q^3 & 0 & 0 & 0 & 0 & 0 & 0 & 0 & 0 \cr
0 & 0 & 0 & 0 & 0 & q^3 & 0 & 0 & 0 & 0 & 0 & 0 & 0 \cr
0 & 0 & 0 & 0 & 0 & 0 & q^{-4} & 0 & 0 & 0 & 0 & 0 & 0 \cr
0 & 0 & 0 & 0 & 0 & 0 & 0 & q^3 & 0 & 0 & 0 & 0 & 0 \cr
0 & 0 & 0 & 0 & 0 & 0 & 0 & 0 & q^{-4} & 0 & 0 & 0 & 0 \cr
0 & 0 & 0 & 0 & 0 & 0 & 0 & 0 & 0 & q^3 & 0 & 0 & 0 \cr
0 & 0 & 0 & 0 & 0 & 0 & 0 & 0 & 0 & 0 & q^3 & 0 & 0 \cr
0 & 0 & 0 & 0 & 0 & 0 & 0 & 0 & 0 & 0 & 0 & q^{-4} & 0 \cr
0 & 0 & 0 & 0 & 0 & 0 & 0 & 0 & 0 & 0 & 0 & 0 & q^3
\end{pmatrix}\ ,\nn
\ea

\medskip

\ba
\pi^{(8)}(b_2) = \begin{pmatrix}
B^0_{~0} & B^0_{~1} & 0 & 0 & 0 & 0 & 0 & 0 & 0 & 0 & 0 & 0 & 0 \cr
B^1_{~0} & B^1_{~1} & 0 & 0 & 0 & 0 & 0 & 0 & 0 & 0 & 0 & 0 & 0 \cr
0 & 0 & q^3 & 0 & 0 & 0 & 0 & 0 & 0 & 0 & 0 & 0 & 0 \cr
0 & 0 & 0 & B^0_{~0}& B^0_{~1} & 0 & 0 & 0 & 0 & 0 & 0 & 0 & 0 \cr
0 & 0 & 0 & B^1_{~0} & B^1_{~1} & 0 & 0 & 0 & 0 & 0 & 0 & 0 & 0 \cr
0 & 0 & 0 & 0 & 0 & q^3 & 0 & 0 & 0 & 0 & 0 & 0 & 0 \cr
0 & 0 & 0 & 0 & 0 & 0 & B^0_{~0} & B^0_{~1} & 0 & 0 & 0 & 0 & 0 \cr
0 & 0 & 0 & 0 & 0 & 0 & B^1_{~0} & B^1_{~1} & 0 & 0 & 0 & 0 & 0 \cr
0 & 0 & 0 & 0 & 0 & 0 & 0 & 0 & B^0_{~0} & B^0_{~1} & 0 & 0 & 0 \cr
0 & 0 & 0 & 0 & 0 & 0 & 0 & 0 & B^1_{~0} & B^1_{~1} & 0 & 0 & 0 \cr
0 & 0 & 0 & 0 & 0 & 0 & 0 & 0 & 0 & 0 & q^3 & 0 & 0\cr
0 & 0 & 0 & 0 & 0 & 0 & 0 & 0 & 0 & 0 & 0 & B^0_{~0} & B^0_{~1} \cr
0 & 0 & 0 & 0 & 0 & 0 & 0 & 0 & 0 & 0 & 0 & B^1_{~0} & B^1_{~1}
\end{pmatrix}\ ,\nn
\ea

\medskip

\ba
\pi^{(8)}(b_3) = \begin{pmatrix}
q^{-4} & 0 & 0 & 0 & 0 & 0 & 0 & 0 & 0 & 0 & 0 & 0 & 0 \cr
0 & q^3 & 0 & 0 & 0 & 0 & 0 & 0 & 0 & 0 & 0 & 0 & 0 \cr
0 & 0 & B^0_{~0} & 0 & B^0_{~1} & 0 & 0 & 0 & 0 & 0 & 0 & 0 & 0 \cr
0 & 0 & 0 & q^3 & 0 & 0 & 0 & 0 & 0 & 0 & 0 & 0 & 0\cr
0 & 0 & B^1_{~0} & 0 & B^1_{~1} & 0 & 0 & 0 & 0 & 0 & 0 & 0 & 0 \cr
0 & 0 & 0 & 0 & 0 & B^0_{~0} & 0 & B^0_{~1} & 0 & 0 & 0 & 0 & 0 \cr
0 & 0 & 0 & 0 & 0 & 0 & q^3 & 0 & 0 & 0 & 0 & 0 & 0 \cr
0 & 0 & 0 & 0 & 0 & B^1_{~0} & 0 & B^1_{~1} & 0 & 0 & 0 & 0 & 0 \cr
0 & 0 & 0 & 0 & 0 & 0 & 0 & 0 & q^{-4} & 0 & 0 & 0 & 0 \cr
0 & 0 & 0 & 0 & 0 & 0 & 0 & 0 & 0 & q^3 & 0 & 0 & 0 \cr
0 & 0 & 0 & 0 & 0 & 0 & 0 & 0 & 0 & 0 & B^0_{~0} & 0 & B^0_{~1} \cr
0 & 0 & 0 & 0 & 0 & 0 & 0 & 0 & 0 & 0 & 0 & q^3 & 0 \cr
0 & 0 & 0 & 0 & 0 & 0 & 0 & 0 & 0 & 0 & B^1_{~0} & 0 & B^1_{~1}
\end{pmatrix}\ ,\nn
\ea

\medskip

\ba
\pi^{(8)}(b_4) = \begin{pmatrix}
B^0_{~0} & 0 & 0 & B^0_{~1} & 0 & 0 & 0 & 0 & 0 & 0 & 0 & 0 & 0 \cr
0 & B^0_{~0} & 0 & 0 & B^0_{~1} & 0 & 0 & 0 & 0 & 0 & 0 & 0 & 0 \cr
0 & 0 & q^3 & 0 & 0 & 0 & 0 & 0 & 0 & 0 & 0 & 0 & 0 \cr
B^1_{~0} & 0 & 0 & B^1_{~1} & 0 & 0 & 0 & 0 & 0 & 0 & 0 & 0 & 0\cr
0 & B^1_{~0} & 0 & 0 & B^1_{~1} & 0 & 0 & 0 & 0 & 0 & 0 & 0 & 0 \cr
0 & 0 & 0 & 0 & 0 & q^{-4} & 0 & 0 & 0 & 0 & 0 & 0 & 0 \cr
0 & 0 & 0 & 0 & 0 & 0 & q^3 & 0 & 0 & 0 & 0 & 0 & 0 \cr
0 & 0 & 0 & 0 & 0 & 0 & 0 & q^3 & 0 & 0 & 0 & 0 & 0 \cr
0 & 0 & 0 & 0 & 0 & 0 & 0 & 0 & B^0_{~0} & 0 & 0 & B^0_{~1} & 0 \cr
0 & 0 & 0 & 0 & 0 & 0 & 0 & 0 & 0 & B^0_{~0} & 0 & 0 & B^0_{~1} \cr
0 & 0 & 0 & 0 & 0 & 0 & 0 & 0 & 0 & 0 & q^3 & 0 & 0 \cr
0 & 0 & 0 & 0 & 0 & 0 & 0 & 0 & B^1_{~0} & 0 & 0 & B^1_{~1} & 0 \cr
0 & 0 & 0 & 0 & 0 & 0 & 0 & 0 & 0 & B^1_{~0} & 0 & 0 & B^1_{~1}
\end{pmatrix}\ ,\nn
\ea

\medskip

\ba
\pi^{(8)}(b_5) = \begin{pmatrix}
q^{-4} & 0 & 0 & 0 & 0 & 0 & 0 & 0 & 0 & 0 & 0 & 0 & 0 \cr
0 & q^{-4} & 0 & 0 & 0 & 0 & 0 & 0 & 0 & 0 & 0 & 0 & 0 \cr
0 & 0 & q^3 & 0 & 0 & 0 & 0 & 0 & 0 & 0 & 0 & 0 & 0 \cr
0 & 0 & 0 & q^3 & 0 & 0 & 0 & 0 & 0 & 0 & 0 & 0 & 0 \cr
0 & 0 & 0 & 0 & q^3 & 0 & 0 & 0 & 0 & 0 & 0 & 0 & 0 \cr
0 & 0 & 0 & 0 & 0 & B^0_{~0} & 0 & 0 & 0 & 0 & B^0_{~1} & 0 & 0 \cr
0 & 0 & 0 & 0 & 0 & 0 & B^0_{~0} & 0 & 0 & 0 & 0 & B^0_{~1} & 0 \cr
0 & 0 & 0 & 0 & 0 & 0 & 0 & B^0_{~0} & 0 & 0 & 0 & 0 & B^0_{~1} \cr
0 & 0 & 0 & 0 & 0 & 0 & 0 & 0 & q^3 & 0 & 0 & 0 & 0 \cr
0 & 0 & 0 & 0 & 0 & 0 & 0 & 0 & 0 & q^3 & 0 & 0 & 0 \cr
0 & 0 & 0 & 0 & 0 & B^1_{~0} & 0 & 0 & 0 & 0 & B^1_{~1} & 0 & 0 \cr
0 & 0 & 0 & 0 & 0 & 0 & B^1_{~0} & 0 & 0 & 0 & 0 & B^1_{~1} & 0 \cr
0 & 0 & 0 & 0 & 0 & 0 & 0 & B^1_{~0} & 0 & 0 & 0 & 0 & B^1_{~1}
\end{pmatrix}\ ,\nn
\ea

\medskip

\ba
\pi^{(8)}(b_6) = \begin{pmatrix}
B^0_{~0} & 0 & 0 & 0 & 0 & 0 & 0 & 0 & B^0_{~1} & 0 & 0 & 0 & 0 \cr
0 & B^0_{~0} & 0 & 0 & 0 & 0 & 0 & 0 & 0 & B^0_{~1} & 0 & 0 & 0 \cr
0 & 0 & B^0_{~0} & 0 & 0 & 0 & 0 & 0 & 0 & 0 & B^0_{~1} & 0 & 0 \cr
0 & 0 & 0 & B^0_{~0} & 0 & 0 & 0 & 0 & 0 & 0 & 0 & B^0_{~1} & 0 \cr
0 & 0 & 0 & 0 & B^0_{~0} & 0 & 0 & 0 & 0 & 0 & 0 & 0 & B^0_{~1} \cr
0 & 0 & 0 & 0 & 0 & q^3 & 0 & 0 & 0 & 0 & 0 & 0 & 0 \cr
0 & 0 & 0 & 0 & 0 & 0 & q^3 & 0 & 0 & 0 & 0 & 0 & 0 \cr
0 & 0 & 0 & 0 & 0 & 0 & 0 & q^3 & 0 & 0 & 0 & 0 & 0 \cr
B^1_{~0} & 0 & 0 & 0 & 0 & 0 & 0 & 0 & B^1_{~1} & 0 & 0 & 0 & 0 \cr
0 & B^1_{~0} & 0 & 0 & 0 & 0 & 0 & 0 & 0 & B^1_{~1} & 0 & 0 & 0 \cr
0 & 0 & B^1_{~0} & 0 & 0 & 0 & 0 & 0 & 0 & 0 & B^1_{~1} & 0 & 0 \cr
0 & 0 & 0 & B^1_{~0} & 0 & 0 & 0 & 0 & 0 & 0 & 0 & B^1_{~1} & 0 \cr
0 & 0 & 0 & 0 & B^1_{~0} & 0 & 0 & 0 & 0 & 0 & 0 & 0 & B^1_{~1}
\end{pmatrix}\ ,\nn
\ea

\medskip

\ba
\pi^{(8)}(b_7) = \begin{pmatrix}
q^{-4} & 0 & 0 & 0 & 0 & 0 & 0 & 0 & 0 & 0 & 0 & 0 & 0 \cr
0 & q^{-4} & 0 & 0 & 0 & 0 & 0 & 0 & 0 & 0 & 0 & 0 & 0 \cr
0 & 0 & q^{-4} & 0 & 0 & 0 & 0 & 0 & 0 & 0 & 0 & 0 & 0 \cr
0 & 0 & 0 & q^{-4} & 0 & 0 & 0 & 0 & 0 & 0 & 0 & 0 & 0 \cr
0 & 0 & 0 & 0 & q^{-4} & 0 & 0 & 0 & 0 & 0 & 0 & 0 & 0 \cr
0 & 0 & 0 & 0 & 0 & q^3 & 0 & 0 & 0 & 0 & 0 & 0 & 0 \cr
0 & 0 & 0 & 0 & 0 & 0 & q^3 & 0 & 0 & 0 & 0 & 0 & 0 \cr
0 & 0 & 0 & 0 & 0 & 0 & 0 & q^3 & 0 & 0 & 0 & 0 & 0 \cr
0 & 0 & 0 & 0 & 0 & 0 & 0 & 0 & q^3 & 0 & 0 & 0 & 0 \cr
0 & 0 & 0 & 0 & 0 & 0 & 0 & 0 & 0 & q^3 & 0 & 0 & 0 \cr
0 & 0 & 0 & 0 & 0 & 0 & 0 & 0 & 0 & 0 & q^3 & 0 & 0 \cr
0 & 0 & 0 & 0 & 0 & 0 & 0 & 0 & 0 & 0 & 0 & q^3 & 0 \cr
0 & 0 & 0 & 0 & 0 & 0 & 0 & 0 & 0 & 0 & 0 & 0 & q^3
\end{pmatrix}\ .\lb{B8pt}
\ea

\smallskip

\section{The braid group ${\cal B}_n\,$ generators for general $n$}

\setcounter{equation}{0}
\renewcommand\theequation{\thesection.\arabic{equation}}

The explicit form of the Artin braid group ${\cal B}_n\,$ generators for small $n\,$ displayed in the previous section and the recursive construction of the corresponding monodromy representations suggest the following structure for general $n\,.$

The construction of the representation spaces (\ref{Vn}) and of their bases implies that the (irreducible) $d_n$-dimensional representation     $\pi^{(n)}\,$ of ${\cal B}_n\,$ on the space $V_{d_n}\,$ reduces, when restricted to the subgroup ${\cal B}_{n-2} \,$ (cf. (\ref{Bn-incl})), to the direct sum
\ba
\pi^{(n)} ({\cal B}_{n-2}) &=&
\pi^{(n-2)} ({\cal B}_{n-2}) \oplus \pi^{(n-1)} ({\cal B}_{n-2})\ ,\quad
n\ge 4\ , \nn\\
\lb{oplus}\\
\pi^{(n)} (b_i) &=& \pi^{(n-2)} (b_i)\oplus\, \pi^{(n-1)} (b_i)\ ,\quad i = 1, \dots, n-3\ .\nn
\ea
The latter formula is exemplified for $b_1$ in the rudimental case (\ref{n4-12}), for $b_1$ and $b_2$ in (\ref{B5}), for $b_1 , b_2$ and $b_3$ in (\ref{B6}), etc. Applying it for $\pi^{(n-1)} (b_i)\,$ (excluding $b_{n-3}\,$) we obtain
\be
\pi^{(n)} (b_i) = \begin{pmatrix}
\pi^{(n-2)} (b_i) & {\mathbf 0} & {\mathbf 0} \cr
{\mathbf 0} & \pi^{(n-3)} (b_i) & {\mathbf 0} \cr
{\mathbf 0} & {\mathbf 0} & \pi^{(n-2)} (b_i)
\end{pmatrix}\ ,\quad i = 1, \dots, n-4\ .
\lb{oplus+}
\ee

As for the matrices of the last two Artin generators, $\pi^{(n)}(b_{n-2})\,$ and $\pi^{(n)}(b_{n-1})\,$ for $n\ge 5\,,$ the obtained explicit results suggest the following block structure:
\ba
\pi^{(n)}(b_{n-2}) &=&
\begin{pmatrix}
B^0_{~0}\, \id_{d_{n-2}} & {\mathbf 0} & B^0_{~1}\, \id_{d_{n-2}} \cr
{\mathbf 0} & q^3\, \id_{d_{n-3}} & {\mathbf 0} \cr
B^1_{~0}\, \id_{d_{n-2}} & {\mathbf 0} & B^1_{~1} \, \id_{d_{n-2}}
\end{pmatrix}\ ,\nn\\
\lb{Bn-last2}\\
\pi^{(n)}(b_{n-1}) &=&
\begin{pmatrix}
q^{-4} \, \id_{d_{n-2}} & {\mathbf 0} \cr
{\mathbf 0} & q^3 \, \id_{d_{n-1}}\cr
\end{pmatrix} \equiv \begin{pmatrix}
q^{-4} \, \id_{d_{n-2}} & {\mathbf 0} & {\mathbf 0} \cr 
{\mathbf 0} & q^3 \, \id_{d_{n-3}} & {\mathbf 0}\cr
{\mathbf 0} & {\mathbf 0} & q^3 \, \id_{d_{n-2}}
\end{pmatrix} \ .
\nn
\ea
As we have already mentioned, the triple Artin relations in (\ref{Artin}) imply that the determinants of all $\pi^{(n)}(b_i)\,$ in a given representation are equal. Denoting their common value by $D_n\,,$ we obtain from (\ref{oplus}) and (\ref{Bn-last2}) that
\be
D_n = D_{n-2}\, .\, D_{n-1}\ ,\quad D_n = q^{- 4\, d_{n-2} + 3 \, d_{n-1}} =  q^{3\, ( 2\,d_{n-2} + d_{n-1}) } = q^{3\, (2\, d_n - d_{n-1}) }
\lb{det-n}
\ee
where we have used that $q^{-4} = q^{6}\,$ and $d_{n-2} + d_{n-1} = d_n\,.$
Using the recursion with $D_2 = q^{-4}\,,\ D_3 = q^3\,$ or alternatively, the (shifted) Fibonacci number sequence
\ba
d_2 &=& 1\,,\ d_3 = 1\,,\ d_4 = 2\,,\ d_5 = 3\,,\ d_6 = 5\,,\ d_7 = 8\,,\ d_8 = 13\,,\ d_9 = 21\,,\nn\\
d_{10} &=& 34\,,\ \, d_{11} = 55\,,\ \, d_{12} = 89\,,\ \, d_{13} = 144\,,\ \, d_{14} = 233\,,\ \, d_{15} = 377\,,\nn\\
d_{16} &=& 610\,,\ \,  d_{17} = 987\,,\ \, d_{18} = 1597\,,\,\ d_{19} = 2584\,,\,\ d_{20} = 4181\,,\ \dots
\lb{Fibo}
\ea
we obtain
\ba
D_2 &=& q^{-4}\,,\ D_3 = q^3\,,\
D_4 = q^{-1}\,,\ D_5 = q^2\,,\  D_6 = q\,,\ D_7 = q^3\,,\nn\\
D_8 &=& q^4\,,\ D_9 = q^{-3}\,,\ D_{10} = q\,,\ D_{11} = q^{-2}\,,\ D_{12} = q^{-1}\,,\ D_{13} = q^{-3}\,,\nn\\
D_{14} &=& q^{-4}\,,\,\ D_{15} = q^3\,,\ \dots \qquad\quad
(\, D_n = D^{-1}_{n + 6} =  D_{n + 12} \, )\ .
\lb{det-12}
\ea

\smallskip

\section{Computational vectors and qubits}

\setcounter{equation}{0}
\renewcommand\theequation{\thesection.\arabic{equation}}

We are now fully prepared for the final step -- to define the ${\cal N}\,$ qubit spaces and the "computational vectors" forming their bases.

To this end, we will start with the general expression for the $n$-point Fibonacci anyon conformal blocks (\ref{1}) (for $n\,$ even), with the following two simplifications.  First, as the presence of the $3\, r \,$ electrons does not affect the braiding properties, we will just set $r = 0\,.$ We will also get rid at this stage of the redundant first and last pairs of indices ($0 1\,$ and $1 0\,,$ respectively) in (\ref{1}). 

\begin{figure}[htb]
\centering
\includegraphics[width=.6\textwidth]{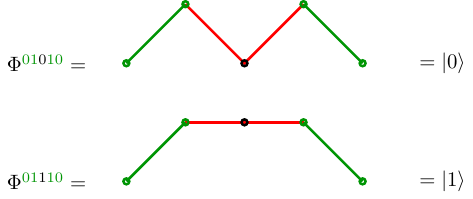}%,trim=0 380 0 200,clip]{Fig1.jpg}
\hfill
\caption{\label{Fig1}{ ${\cal N}=1\,,\ 4\,$ anyons, $\dim\, V_{d_4} = 2\,$} }
\end{figure}

It is obvious from (\ref{n4-23}) that, by a 
slight modification the prescription given in \cite{BHZS05}, a basis in the single qubit (${\cal N}=1\,$) space is given by the two $4$-point Fibonacci conformal blocks ($4 = 2\,.\,1 + 2\,$). To convey the idea suitable for generalization to higher ${\cal N}\,$ we will introduce below both a graphical notation (in terms of Bratteli path diagrams, see Figure~\ref{Fig1}),
and an equivalent shorthand notation,
\be
\begin{pmatrix}
\Phi^{{\textcolor{deepgreen}{01}}0{\textcolor{deepgreen}{10}}}\cr
\Phi^{{\textcolor{deepgreen}{01}}1{\textcolor{deepgreen}{10}}}
\end{pmatrix} 
=
\begin{pmatrix}
\textcolor{deepgreen}{\langle \e |}\, (\e\, _0\, \e ) \,\textcolor{deepgreen}{ | \e \rangle}\cr
\textcolor{deepgreen}{\langle \e |}\, (\e\, _1\, \e ) \,\textcolor{deepgreen}{ | \e \rangle}
\end{pmatrix} 
= :
\begin{pmatrix}
| 0 \,\rangle\cr
| 1 \,\rangle
\end{pmatrix} \ ,
\lb{eae}
\ee
where 
$(\e\, _\a\, \e )\,,\ \a = 0,1\,$ stays, in general, for $\Pi_1 \, \e (w_{i+1})\, \Pi_\a\, \e (w_{i+2})\, \Pi_1\,$ and 
\begin{equation*}
\langle \e | = \langle 0 | \,\e (w_1)\,\Pi_1\ ,\qquad 
|\e \rangle = \Pi_1\,\e (w_{2\,{\cal N}+2})\,|0\rangle\ .
\end{equation*}
We will call in what follows the configurations in red depicted in Figure~\ref{Fig1} "(triangular) pails" and "double ropes", respectively. The first and the last $\e\,$ fields in green are actually {\it inert} in the sense that, intertwining between the (left, resp. the right) vacuum sector and the non-trivial one, they perform a {\it single channel} map in both cases.

\medskip

\noindent
{\bf Remark 5~} The correspondence between the qubit space construction in terms of Fibonacci $4$-point conformal blocks and its $3$-quasiparticle (of $q$-spin $1$) counterpart in \cite{BHZS05} goes as follows. The two "computational states" depicted in Fig. $1\,$ of \cite{BHZS05} are actually realized directly by the configurations of the first three anyons in {\it our} Figure~\ref{Fig1} above. In the quantum group picture, the fusion rules (\ref{01}) are equivalent to the tensor product decomposition of quantum spins $0\,$ and $1\,$ at $q^5 = -1\,,$ the $q$-spin $I\,$ representations being characterized by their {\it quantum dimensions} 
\be
[I]\ ,\ \ I = 0,1 \qquad \rightarrow \qquad [2I + 1]_q := \frac{q^{2I+1} - q^{-(2I+1)}}{q-q^{-1}}\ .
\lb{qn}
\ee 
For $q=e^{i\frac{\pi}{5}}\,,$ the relevant $q$-numbers appearing in the $q$-analog of the Clebsch-Gordan decomposition are
\be
[1]_q = 1\ ,\qquad [3]_q = q^2 + 1 + q^{-2} = \tau + 1 = \tau^{-1}\ ,\qquad [5]_q = 0
\lb{qI}
\ee
(cf. (\ref{tau})). The "truncation" (the lack of $q$-spin $2\,,$ in this case) in the nontrivial fusion relation $[1 ] \times [1 ] = [0 ] \oplus [1 ]\,$ in (\ref{01}) is reflected in an identity arising from the expansion
\be
[3]_q \,.\, [3]_q = [1]_q + [3]_q + [5]_q \qquad \Rightarrow\qquad \tau^{-2} = 1 + \tau^{-1} + 0\ .
\lb{q1}
\ee

In the $4$-point conformal block realization of the qubit space, the braiding of any two neighboring pairs of Fibonacci anyons provides a representation of ${\cal B}_4\,$ generated by (\ref{n4-12}) and (\ref{n4-23})
\ba
&&\pi^{4} (b_1) \,|\a\rangle = 
R^\a_{~\b}\, |\b\rangle = 
\pi^{4} (b_3) \,|\a\rangle\ ,\qquad 
R^\a_{~\b} = q^{3(2-\a)}\, \d^\a_\b\ ,\nn\\
&&\pi^{4} (b_2) \,|\a\rangle = 
B^\a_{~\b}\,|\b\rangle\ ,\qquad \a , \b = 0 , 1
\lb{RBa}
\ea
(summation over $\b\,$ is assumed). Due to the presence in (\ref{RBa}) of the non-diagonal $2\times 2\,$ matrix $B = (B^\a_{~\b})\,,$ (\ref{n4-23}) provides the simplest non-Abelian braid group representation needed for (topological) quantum computation.

It should be noted that the three anyon "non-computational (NC)" state of \cite{BHZS05} does 
not appear in this realization which is quite welcome, since "non-computational" actually means  "redundant". In a sense, the counterpart of the NC state is a $3$-point function on which the braiding  acts simply as multiplication by $q^3\,,$ cf. (\ref{n23}). 

The generalization to ${\cal N}\,$ qubits looks now straightforward. We will define the ${\cal N}\,$ qubit {\it computational states} as
\ba 
&&|\,\a_1\, \a_2\, \dots\, \a_{\cal N} \,\rangle := 
\textcolor{deepgreen}{\langle \e |}\,  (\e\, _{\a_1}\,\e) \,  (\e\, _{\a_2}\,\e) \, \dots\, (\e\, _{\a_{\cal N}}\,\e) \, \textcolor{deepgreen}{| \e \rangle} = \nn\\
&&= \Phi^{\textcolor{deepgreen}{0\, 1}\, \a_1 \, 1 \, \a_2 \,1\, \dots \, 1\, \a_{{\cal N}} \,\textcolor{deepgreen}{1\, 0 }} = \textcolor{deepgreen}{\langle \e |}\,  \e (w_2)\,\Pi_{\a_1}\, \e (w_3)\,\Pi_1\,\dots \, \Pi_1\,\e (w_{2\,{\cal N}})\, \Pi_{\a_{\cal N}}\, \e (w_{2\,{\cal N}+1})\, \textcolor{deepgreen}{| \e \rangle}\qquad
\lb{4}
\ea
(the rule is that every second projector in (\ref{4}) is $\Pi_1\,$), for $i = 1,\dots, {\cal N}\,$ and $\a_i = 0,1\,.$ 
Graphically, the Bratteli diagrams of even total length corresponding to computational states only contain configurations of  triangular pails, for $\a_i = 0\,$ and double ropes, for $\a_i = 1\,.$ The rest of the conformal blocks, i.e. those containing also ropes of {\it odd} number of segments\footnote{The total number of such segments has to be even, of course.} correspond to NC states. 

\begin{figure}[tbp]
\centering
\includegraphics[width=\textwidth]{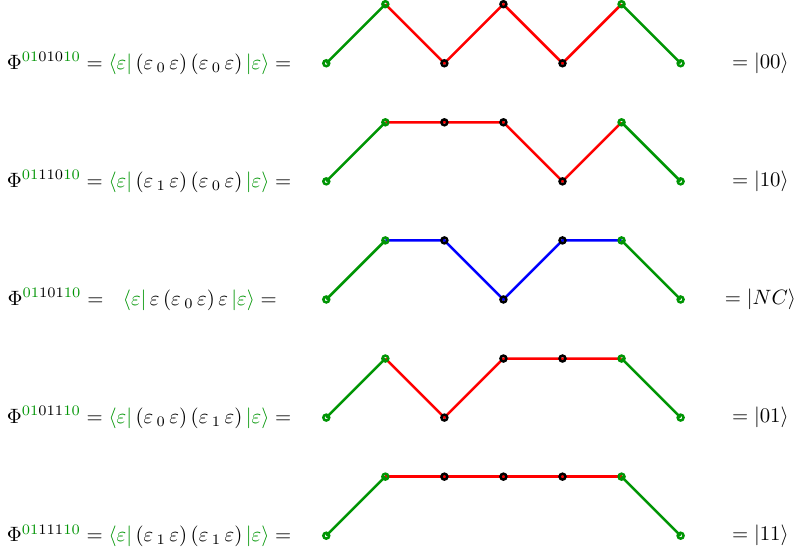}%,trim=0 380 0 200,clip]{Fig2.jpg}
\hfill
\caption{\label{Fig2}{ ${\cal N}=2\,,\ 6\,$ anyons, $\dim\, V_{d_6} = 5\,, \ 2$-qubit computational vectors in red, NC in blue}  }
\end{figure}

\begin{figure}[tbp]
\centering
\includegraphics[width=0.73\textwidth]{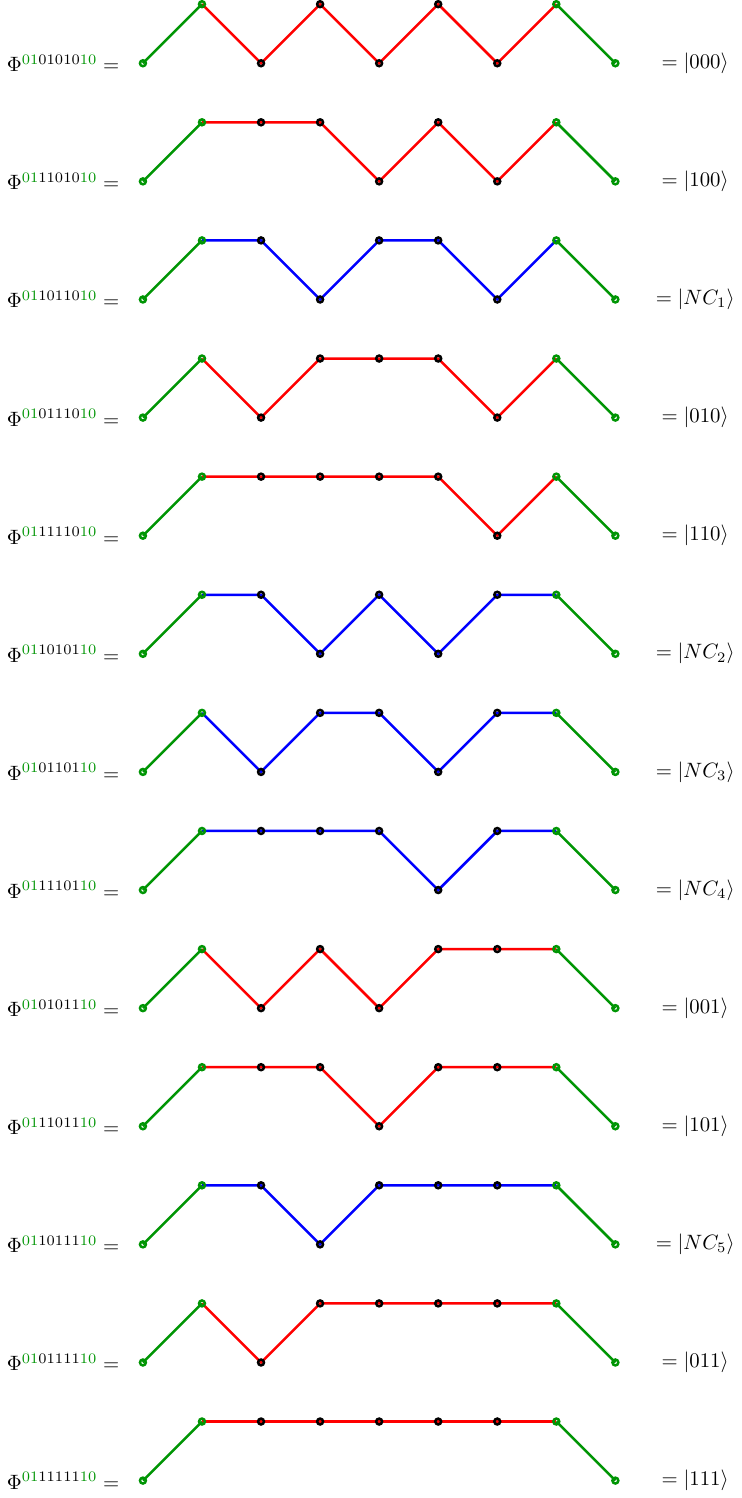}
%,trim=0 380 0 200,clip]{Fig3.jpg}
\hfill
\caption{\label{Fig3}{${\cal N}=3\,,\ 8\,$ anyons, $\dim\, V_{d_8} = 13\,, \ 3$-qubit computational vectors in red, NC ones in blue}  }
\end{figure}

We proceed with the ${\cal N}=2\,$ and ${\cal N}=3\,$ examples, cf. Figures~\ref{Fig2} and \ref{Fig3}.

Obviously, when ${\cal N}\,$ is increased by $1\,$ the number of computational states is doubled, starting with $2\,$ for ${\cal N}=1\,,$ so it is equal to $2^{\cal N}\,$ as it should. The number of the rest (NC) vectors is, accordingly, $d_{2({\cal N}+1)} - 2^{\cal N}\,,$ i.e. $0\,, 1\,, 5\,, \dots\,$ for ${\cal N} = 1\,, 2\,, 3\,, \dots\,,$ see (\ref{Fibo}). As $d_{2\,{\cal N}} > 2^{\cal N}\,$ for ${\cal N} \ge 5\,,$ the number of NC states exceeds that of the computational ones already for $10\,$ anyons.

To justify the validity of the realization described above as a genuine ${\cal N}\,$ qubit space for $2\,{\cal N}+2\,$ anyons (${\cal N}\ge 2$), we will show that no leakage occurs, i.e. there is no mixing of the braid group action on computational and NC states. To this end, we must specify a subgroup of ${\cal B}_{2\,{\cal N}+2}\,$ which preserves the arrangement of the ${\cal N}\,$ qubits and in the same time, still provides the needed quantum computational tools (including the Solovay-Kitaev algorithm, see e.g. \cite{HZBS07} for details). Such a subgroup is the one generated by $b_1\,,\, b_{2i}\,,\ i = 1, 2, \dots, {\cal N}\,$ and $b_{2\,{\cal N}+1}\,,$ i.e. the first, the last and all even generators of ${\cal B}_{2\,{\cal N}+2}\,,$ having the direct product structure
\be
{\cal B}^{({\cal N})}_{sub} = {\cal B}_3 \times ({\cal B}_2)^{\times ({\cal N}-2)}  \times{\cal B}_3 \equiv {\cal B}_3 \times {\cal B}_2 \times \dots \times 
{\cal B}_2 \times {\cal B}_3 \subset 
{\cal B}_{2\,{\cal N}+2}\ ,\qquad {\cal N}\ge 2\ .
\lb{DPB}
\ee
Note that the even generators commute with each other as a consequence of the Artin relations (\ref{BArt}). The action of the group (\ref{DPB}) on the Fibonacci conformal block space $V_{d_{2\,{\cal N}+2}}\,$ will be displayed below. This is not a problem as it reduces, essentially, to writing down (only) part of the matrices derived in the previous sections; the actually important thing will be to reveal the structure of the corresponding representation and get convinced, if this is the case, that it is fully reducible (i.e., decomposable), being a direct sum of the computational and NC spaces.

We will start, for completeness, with the ${\cal N}=1\,$ case ($4\,$ anyons) where no NC states are present and no restriction is needed, writing once more the full ${\cal B}_4\,$ action displayed in (\ref{RBa}) in the form
\be
b_1\, | \a \rangle = R^\a_{~\b}\, | \b \rangle\ ,\qquad
b_2\, | \a \rangle = B^\a_{~\b}\, | \b \rangle\ ,\qquad
b_3\, | \a \rangle = R^\a_{~\b}\, | \b \rangle\ .\quad
\lb{BresN1}
\ee
For ${\cal N}=2\,$ ($6\,$ anyons) we obtain from (\ref{B6}) and Fig. 2
\ba
b_1\, | \a_1 \,\a_2 \rangle &=& R^{\a_1}_{~{\b}_1} \d^{\a_2}_{\b_2}\, | \b_1 \,\b_2 \rangle\ ,\qquad
b_2\, | \a_1 \,\a_2 \rangle = B^{\a_1}_{~{\b}_1} \d^{\a_2}_{\b_2}\, | \b_1 \,\b_2 \rangle\ ,\nn\\
b_4\, | \a_1\, \a_2 \rangle &=& \d^{\a_1}_{\b_1}\,B^{\a_2}_{~{\b}_2} \, | \b_1 \,\b_2 \rangle\ ,\qquad
b_5\, | \a_1\, \a_2 \rangle = \d^{\a_1}_{\b_1}\,R^{\a_2}_{~{\b}_2}\,  | \b_1 \,\b_2 \rangle\ ,\nn\\
b_i\, | NC \rangle &=& q^3 \,| NC \rangle\ ,\quad i = 1,2,4,5\ .
\lb{BresN2}
\ea 
We will write down once more the diagonal action of $b_1\,$ and $b_5\,$ in the form
\ba
b_1\, | 0\, 0 \rangle &=& q^{-4}\, | 0\, 0 \rangle \ ,\quad 
b_1\, | 1 \,0 \rangle = q^3\, | 1\, 0 \rangle \ ,\quad\ \,
b_1\, | 0\, 1 \rangle = q^{-4}\, | 0\, 1 \rangle \ ,\quad 
b_1\, | 1 1 \rangle = q^3\, | 1 1 \rangle \ ,\nn\\
b_5\, | 0\, 0 \rangle &=& q^{-4}\, | 0\, 0 \rangle \ ,\quad 
b_5\, | 1\, 0 \rangle = q^{-4}\, | 1\, 0 \rangle \ ,\quad 
b_5\, | 0\, 1 \rangle = q^3\, | 0\, 1 \rangle \ ,\quad\ \, 
b_5\, | 1 1 \rangle = q^3\, | 1 1 \rangle\ .\qquad\qquad
\lb{diag}
\ea
The result displayed in (\ref{BresN2}) is very encouraging. It shows that the representation of the group ${\cal B}_{sub}^{(2)} = {\cal B}_3 \times {\cal B}_3\,$ on the $5\,$ dimensional space $V_{d_6}\,$ (\ref{6pt1}) is indeed fully reducible so that the $4 = 2^2\,$ computational vectors split from the NC one, forming a direct sum $V_{d_6} = V_{2^2} \oplus V_{NC}\,.$ Further, the $2\,$ qubit space $V_{2^2}\,$ itself has the form of a tensor product of two single qubit spaces on which the ${\cal B}_{sub}^{(2)}\,$ representation is, accordingly, the tensor square of the two single qubit ${\cal B}_3\,$ representations, i.e.
\be
V_{2^2} = V_2 \otimes V_2\ ,\qquad
{\cal B}^{(2)}_{sub} \, V_{2^2} = {\cal B}_3 \, V_2  \otimes {\cal B}_3\, V_2\ ,\qquad 
{\cal B}^{(2)}_{sub} \, V_{NC} = q^3\, V_{NC}\ ,
\lb{Bsub2rep}
\ee
the first ${\cal B}_3\,$ group in (\ref{Bsub2rep}) being generated by $b_1\,$ and $b_2\,$ and the second, by $b_4\,$ and $b_5\,.$

\medskip

\noindent
{\bf Remark 6~} The computational vectors in (\ref{diag}) appear in the order inherited from the ordering of all vectors introduces recursively in Section 5 (see the paragraph after (\ref{Vn}). It does not coincide with the commonly used lexicographical order (that would be $| 00 \rangle\,,\ | 01 \rangle\,,\ | 10 \rangle\,,\ | 11 \rangle$)\footnote{The chosen ordering is actually "colexicographical".}. In any case, writing down  the restrictions of the braid matrices on the ${\cal N}$ qubit space for ${\cal N}=2\,$ in the form (\ref{BresN2}) shows that they are given by the tensor products
\be
b_1 = R \otimes \id\ ,\qquad b_2 = B \otimes \id\ ,\qquad b_4 = \id \otimes B\ ,\qquad b_5 = \id \otimes R
\lb{BresN22}
\ee
expressed in matrix form as Kronecker product, 
$(A\otimes B)^{\a_1 \a_2}_{~\b_1 \b_2} := A^{\a_1}_{~\b_1}\, B^{\a_2}_{~\b_2}\,.$

\medskip

Having the experience with the first non-trivial $2$ qubit case we can return to the  general recursive formulae (\ref{oplus+}) and (\ref{Bn-last2}) implying
\ba
\pi^{(2\,{\cal N}+2)}(b_{i}) &=&
\begin{pmatrix}
\pi^{(2\,{\cal N})} (b_{i}) & {\mathbf 0} & {\mathbf 0} \cr
{\mathbf 0} & \pi^{(2\,{\cal N}-1)} (b_{i}) & {\mathbf 0} \cr
{\mathbf 0} & {\mathbf 0} & \pi^{(2\,{\cal N})} (b_{i})
\end{pmatrix}\ ,\nn\\ 
{\rm for}\quad i &=& 1 \quad {\rm and}\quad i = 2j\ ,\quad j= 1\,,\,2\,\, \dots\,,\, {\cal N}-1\ ,\nn\\
\pi^{(2\,{\cal N}+2)}(b_{2\,{\cal N}}) &=&
\begin{pmatrix}
B^0_{~0}\, \id_{d_{2\,{\cal N}}} & {\mathbf 0} & B^0_{~1}\, \id_{d_{2\,{\cal N}}} \cr
{\mathbf 0} & q^3\, \id_{d_{2\,{\cal N}-1}} & {\mathbf 0} \cr
B^1_{~0}\, \id_{d_{2\,{\cal N}}} & {\mathbf 0} & B^1_{~1} \, \id_{d_{2\,{\cal N}}}
\end{pmatrix}\ ,
\lb{Bn-last21}\\
\pi^{(2\,{\cal N}+2)}(b_{2\,{\cal N}+1}) &=&
\begin{pmatrix}
q^{-4} \, \id_{d_{2\,{\cal N}}} & {\mathbf 0} & {\mathbf 0} \cr
{\mathbf 0} & q^3 \, \id_{d_{2\,{\cal N}-1}} & {\mathbf 0} \cr
{\mathbf 0} & {\mathbf 0} & q^3 \, \id_{d_{2\,{\cal N}}} 
\end{pmatrix}\ 
\nn
\ea
to find the origin of the NC vectors that "speckle" the list of computational ones. NC states are related to the appearance of the central square block of size $d_{2\,{\cal N}-1}\,,$ for the first time for ${\cal N}=2\,$ ($d_3 = 1$), and proliferate in the subsequent representations with the increase of ${\cal N}\,.$

We recall that, by definition, the ${\cal N}\,$ qubit subspace $V_{2^{\cal N}}\,$ is the ${\cal N}$-th tensor power of the single qubit one:
\be
V_{2^{\cal N}} := V_2^{\otimes {\cal N}} \equiv V_2 \otimes \dots \otimes V_2\quad ({\cal N}\ {\rm times})\ ,\qquad 
|\a_1 \,\a_2\,\dots \, \a_{\cal N}\,\rangle \equiv
|\a_1 \,\rangle \otimes |\a_2\,\,\rangle \dots \, |\a_{\cal N}\,\rangle \ .
\lb{NqV}
\ee
Physically, this means that the individual qubits are independent. 

In the Fibonacci conformal block realization (for even number $2\, {\cal N} + 2\,$ of anyons placed on a one-dimensional boundary, and inert end ones) the separate qubit spaces are realized by braiding the second and the third, the fourth and the fifth etc., till $\,2\, {\cal N}\,$ and $2\, {\cal N} + 1\,.$ The representation of the group (\ref{DPB}) on the full $d_{2\,{\cal N}+2}\,$ dimensional space of conformal blocks is fully reducible, leaving invariant both the 
${\cal N}\,$ qubit subspace (\ref{NqV}) formed by the computational vectors defined above (those whose Bratteli diagrams only contain "triangular pails" and "double ropes") and its linear complement $V_{NC}\,$ spanned by the remaining, non-computational vectors. The latter assertion is the gist of the "no leaking" theorem.

A general proof of the above statement can be carried out by induction, stepping essentially on the block matrix form of $\pi^{(2\,{\cal N}+2)}(b_{2\,{\cal N}})\,$ 
(\ref{Bn-last21}).

\medskip

\noindent
{\bf Remark 7~} The backward iteration of the block structure displayed in (\ref{Bn-last21}) suggests that the shifted Fibonacci numbers (\ref{Fibo}) satisfy the identity
\ba
d_{2\,{\cal N}+2} &=& 2^{{\cal N}-1} \,d_4 + \sum_{k=0}^{{\cal N}-2} 2^{{\cal N}-2-k}\, d_{2k+3}
\equiv 2^{{\cal N}-1} \,d_4 + \sum_{k=0}^{{\cal N}-2} 2^k\, d_{2({\cal N}-k)-1} = \nn\\
&=&  2^{\cal N} + 2^{{\cal N}-2}\,d_3 + 2^{{\cal N}-3}\, d_5 + \dots + 2\,d_{2\,{\cal N}-3} + d_{2\,{\cal N}-1} \ .
\lb{Fib}
\ea
Its proof is elementary; we have
\ba
d_{n+1} &=& d_n + d_{n-1}\ ,\quad n\ge 3\ ,\qquad d_2 = 1 = d_3\qquad\Rightarrow\nn\\
d_{2\,{\cal N}+2} &=& d_{2\,{\cal N}} + d_{2\,{\cal N}+1} = 2\,d_{2\,{\cal N}} + d_{2\,{\cal N}-1}\ ,\quad {\cal N}\ge 1\ ,\qquad 
d_1 = 0\ ,
\lb{Fib-ind}
\ea
etc.

\medskip

Instead of dwelling on the general ${\cal N}\,$ case we would invite the interested reader to verify the following results for ${\cal N}=3\,$ ($8\,$ anyons) using (\ref{B8pt}) and Fig. 3 (there are $5\,$ NC vectors, of $13\,$ altogether in this configuration, on rows with numbers $3, 6, 7, 8\,$ and $11\,$ in Fig. 3): 
\ba
V_{d_8} &=& V_{2^3} \oplus V^{(3)}_{NC}\ ,\qquad V_{2^3} = V_2 \otimes V_2 \otimes V_2\ ,\qquad
\dim\, V_{d_8} = 13\ ,\qquad
\dim\, V^{(3)}_{NC} = 5\ ,\nn\\
{\cal B}^{(3)}_{sub} &=& {\cal B}_3 \times {\cal B}_2 \times {\cal B}_3\ ,\qquad b_1 = R \otimes \id \otimes \id\ ,\qquad b_2 = B \otimes \id \otimes \id\ ,\nn\\
b_4 &=& \id \otimes B \otimes \id\ ,\qquad
b_6 = \id \otimes \id\otimes  B\ , \qquad 
b_7 = \id \otimes \id \otimes R\ .
\lb{BresN33}
\ea
Although not being of central interest, we will also display the action of ${\cal B}^{(3)}_{sub}\,$
on the NC sector. It is also fully reducible, the space $V^{(3)}_{NC}\,$ being decomposed into a 
$2\,$ dimensional invariant subspace spanned by 
$|NC_1\rangle\,$ and $|NC_5\rangle\,,$ and a $3\,$ dimensional one, spanned by $|NC_2\rangle\,, \ |NC_3\rangle\,$ and $|NC_4\rangle\,,$ respectively, so that
\ba
b_1 &=& q^3 \id_2 \oplus \begin{pmatrix} q^3 & {\mathbf 0}\cr {\mathbf 0} & R \end{pmatrix}\ ,\qquad b_2 = q^3 \id_2 \oplus \begin{pmatrix} q^3 & {\mathbf 0}\cr {\mathbf 0} & B \end{pmatrix}\ ,
\qquad b_4 = q^3 \id_2 \oplus \begin{pmatrix} q^{-4} & {\mathbf 0}\cr {\mathbf 0} & q^3 \id_2 \end{pmatrix}\ ,
\nn\\
b_6 &=& B \oplus q^3 \id_3 \ ,\qquad
b_7 = R \oplus q^3 \id_3 \ .
\lb{NC3}
\ea

\smallskip

%%%%%%%%%%%%%%%%%%%%%%%%%%%%%%%%%%%%%%%%%%%%%%%%%%%%%%%%%%%%%%%%%

\acknowledgments
\addcontentsline{toc}{section}{Acknowledgments}

This work has been done under the project BG05M2OP001-1.002-0006 "Quantum Communication, Intelligent Security Systems and Risk Management" (QUASAR) financed by the Bulgarian Operational Programme "Science and Education for Smart Growth" (SESG) co-funded by the ERDF. Both LH and LSG thank the Bulgarian Science Fund for partial support under Contract No. DN 18/3 (2017). LSG has been also supported as a Research Fellow by the Alexander von Humboldt Foundation. 

\smallskip

%%%%%%%%%%%%%%%%%%%%%%%%%%%%%%%%%%%%%%%%%%%%%%%%%%%%%%%%%%%%%%%%%%

\end{document}